\documentclass[11pt]{article}
\usepackage[english]{babel}
\usepackage[utf8]{inputenc}
\usepackage[T1]{fontenc}
\usepackage{amsfonts}
\usepackage{amsmath}
\usepackage{amssymb}
\usepackage{amsthm}
\usepackage{bm}

\newcommand{\A}{\mathbb{A}}
\newcommand{\C}{\mathbb{C}}
\renewcommand{\H}{\mathbb{H}}
\newcommand{\N}{\mathbb{N}}
\newcommand{\R}{\mathbb{R}}

\newcommand{\Z}{\mathbb{Z}}
\newcommand{\bA}{{\bm{A}}}
\newcommand{\bF}{{\bm{F}}}

\newcommand{\boC}{\mathcal{C}}

\newcommand{\boE}{\mathcal{E}}
\newcommand{\boF}{\mathcal{F}}
\newcommand{\boG}{\mathcal{G}}
\newcommand{\boH}{\mathcal{H}}
\newcommand{\boL}{\mathcal{L}}

\newcommand{\boO}{\mathcal{O}}

\newcommand{\boQ}{\mathcal{Q}}
\newcommand{\boR}{\mathcal{R}}

\newcommand{\boU}{\mathcal{U}}

\newcommand{\bsalpha}{\bm{\alpha}}
\newcommand{\bsbeta}{\bm{\beta}}
\newcommand{\bssigma}{\bm{\sigma}}

\newcommand{\gS}{\mathfrak{S}}

\DeclareMathOperator{\curl}{{\rm curl}}
\renewcommand{\div}{\mathop{\mathrm{div}}\nolimits}
\DeclareMathOperator{\tr}{{\rm tr}}

\newtheorem*{propo}{Proposition}
\newtheorem{prop}{Proposition}
\newtheorem*{theo}{Theorem}
\newtheorem{thm}{Theorem}
\theoremstyle{definition}
\newtheorem*{rem}{Remark}

\begin{document}

\title{Two Hartree-Fock models for the vacuum polarization}
\author{
\renewcommand{\thefootnote}{\arabic{footnote}}
Philippe Gravejat\footnotemark[1], Christian Hainzl\footnotemark[2], Mathieu Lewin\footnotemark[3]~ and \'Eric S\'er\'e\footnotemark[4]}
\footnotetext[1]{Centre de Math\'ematiques Laurent Schwartz (UMR 7640), \'Ecole Polytechnique, F-91128 Palaiseau Cedex, France. E-mail: {\tt gravejat@math.polytechnique.fr}}
\footnotetext[2]{Mathematisches Institut, Auf der Morgenstelle 10, D-72076 T\"ubingen, Germany. E-mail: {\tt christian.hainzl@ uni-tuebingen.de}}
\footnotetext[3]{Centre National de la Recherche Scientifique and Laboratoire de Math\'ematiques (UMR 8088), Universit\'e de Cergy-Pontoise, F-95000 Cergy-Pontoise, France. E-mail: {\tt Mathieu.Lewin@math.cnrs.fr}}
\footnotetext[4]{Centre de Recherche en Math\'ematiques de la D\'ecision (UMR 7534), Universit\'e Paris-Dauphine, Place du Mar\'echal De Lattre de Tassigny, F-75775 Paris Cedex 16, France. E-mail: {\tt sere@ceremade.dauphine.fr}}
\maketitle

\begin{abstract}
We review recent results about the derivation and the analysis of two Hartree-Fock-type models for the polarization of vacuum. We pay particular attention to the variational construction of a self-consistent polarized vacuum, and to the physical agreement between our non-perturbative construction and the perturbative description provided by Quantum Electrodynamics.
\end{abstract}

\section{Introduction}
\label{Intro}

During the past century, the classical picture of the vacuum as an empty object was challenged by a series of theoretical advances and experimental observations including the measurement of the Lamb shift \cite{LambRet1} and the derivation of the Casimir effect \cite{Casimir1} (see also \cite{CasiPol1}). The description of the vacuum as a complicated fluctuating system emerged. An intuitive picture for this system can be derived from the observation by Blackett and Occhialini \cite{BlacOcc1} of the creation of electron-positron pairs when one provides a sufficient amount of energy to the vacuum. Since the energy-time uncertainty principle \cite{MandTam1} allows important fluctuations of energy during short time intervals, nothing prevents the vacuum from being the place of permanent creations and annihilations of pairs of virtual particles.

This phenomenon affects the interactions between physical particles. It is in particular at the origin of the vacuum polarization. In presence of an external electromagnetic field, the virtual electron-positron pairs modify the distributions of charges and currents originally generated by the fields. The corrections with respect to the original distributions are computed using Quantum Electrodynamics (see e.g. \cite{PeskSch1, GreiRei0, Dyson0}). This perturbative theory provides their value in terms of a power series with respect to the Sommerfeld fine-structure constant
$$\alpha = \frac{e^2}{4 \pi \varepsilon_0 \hbar c}.$$
Unfortunately, divergences appear at any order in the computations. In order to obtain a well-defined value, one has to appeal to complicated renormalization procedures, which provide corrections in an extremely accurate agreement with physical experiments in spite of their intricacy (see e.g. \cite{GaHaKNO1}).

Our goal in this survey is to describe recent mathematical results concerning simplified models for the vacuum polarization. The main difference with Quantum Electrodynamics lies in the non-perturbative nature of the models. The fine-structure constant $\alpha$ being fixed, the polarized vacuum is constructed by variational arguments. The main difficulty lies in the choice of the approximations to make in order to allow such a construction and to guarantee its relevance with respect to the perturbative computations of Quantum Electrodynamics.

The results in this survey are based on a seminal paper by Chaix and Iracane \cite{ChaiIra1} (see also \cite{ChaIrLi1}), which provides a mean-field framework for the analysis of the vacuum polarization. In this setting, the system under consideration is composed of the physical electrons and positrons, coupled with virtual ones which give account of the polarized vacuum. All of them interact with photons, the interactions being instantaneous. The electrons and positrons are described using an Hartree-Fock approximation. Under this approximation, and using renormalization procedures which are quite standard in the context of Quantum Electrodynamics, it becomes possible to define rigorously an energy for the system. The polarized vacuum is then mathematically constructed as a critical point of this energy. One can compare the expansions of its charge and current densities with respect to the fine-structure constant $\alpha$ to the ones provided by Quantum Electrodynamics, and check, at least in the purely 
electrostatic case, that the approximations made in order to fashion the simplified models are quite reasonable. 

This is in brief what we are going to describe in this survey. In the first section, we focus on the derivation of the two Hartree-Fock models that we are going to analyze mathematically. We provide some further details on the notion of polarized vacuum, as well as on the nature of an Hartree-Fock approximation. The second section is devoted to the mathematical analysis of our first model, the so-called Bogoliubov-Dirac-Fock model, which corresponds to the purely electrostatic case. We give rigorous existence results for the polarized vacuum, and check (in the reduced case) the relevance of our definitions with respect to Quantum Electrodynamics by computing an asymptotic expansion of the total charge density with respect to the fine-structure constant $\alpha$. In the third section, we consider the Pauli-Villars regulated model which takes account of the electromagnetic interactions with the photons. Our main results about this further model deal with its rigorous definition and with the mathematical 
construction of a polarized vacuum.

\section{Derivation of the Hartree-Fock models for the vacuum polarization}
\label{sec:deriv}

\subsection{The picture of the Dirac sea}
\label{sub:mer}

In relativistic quantum mechanics, individual electrons are represented by spinors $\psi \in L^2(\R^3, \linebreak[0] \C^4)$. This description originates in the works of Dirac \cite{Dirac1, Dirac2, Dirac3}, who introduced the formula
$$E_c(\psi) = \langle D_{m, 0}(\psi), \psi \rangle_{L^2(\R^3, \C^4)},$$
for the computation of the kinetic energy of a relativistic electron. The free Dirac operator $D_{m, 0}$ is defined as
$$D_{m, 0} = \hbar c \bsalpha \cdot (- i \nabla) + m c^2 \bsbeta,$$
where $\hbar$, $c$ and $m$ stand respectively for the reduced Planck constant, the speed of light (in the free vacuum), and the (bare) mass of an electron. Without loss of generality, we can make a choice of units such that $\hbar = 1$ and $c = 1$. In the sequel, we adopt this choice, so that we drop the dependence on $\hbar$ and $c$ of the operator $D_{m, 0}$.

The Dirac operator is a self-adjoint operator on $L^2(\R^3, \C^4)$, with domain $H^1(\R^3, \C^4)$ (see e.g. \cite{Thaller1}). The Dirac matrices $\bsalpha = (\bsalpha_1, \bsalpha_2, \bsalpha_3)$ and $\bsbeta$ are given by the formulae
$$\bsalpha_k = \begin{pmatrix} 0 & \bssigma_k \\ \bssigma_k & 0 \end{pmatrix} \quad {\rm and} \quad \bsbeta = \begin{pmatrix} I_2 & 0 \\ 0 & - I_2 \end{pmatrix},$$
where the Pauli matrices $\bssigma_1$, $\bssigma_2$ and $\bssigma_3$ are equal to
$$\bssigma_1 = \begin{pmatrix} 0 & 1 \\ 1 & 0 \end{pmatrix}, \quad \bssigma_2 = \begin{pmatrix} 0 & -i \\ i & 0 \end{pmatrix} \quad {\rm and} \quad \bssigma_3 = \begin{pmatrix} 1 & 0 \\ 0 & - 1 \end{pmatrix}.$$
The Dirac matrices are designed so as to satisfy the identity
$$D_{m, 0}^2 = - \Delta + m ^2 I.$$
As a result, the spectrum of the free Dirac operator splits into two intervals according to the expression
$$\sigma(D_{m, 0}) = (- \infty, - m] \cup [m, + \infty).$$
The spectrum is interpreted as the possible levels of kinetic energy for a relativistic electron, so that nothing prevents the kinetic energies to be arbitrarily negative. Such a phenomenon has never been observed in practice.

Dirac \cite{Dirac1, Dirac2, Dirac3} by-passed the difficulty introducing the picture of the Dirac sea. In the free vacuum, an infinite number of virtual electrons completely fill in the levels of negative energy. The possible levels of energy for ``physical'' electrons are positive. The picture amounts to claiming that the free vacuum is not represented by a vanishing quantity, but instead is identified to the negative spectrum of the free Dirac operator. In the following where we will consider electrons in an Hartree-Fock state (see Subsection \ref{sub:Hartree}), the free vacuum will be more precisely identified to the negative spectral projector
$$P_{m, 0}^- = \chi_{( - \infty, 0]}(D_{m, 0}).$$

In presence of an additional external electromagnetic four-potential
$$\bA_{\rm ext} = (V_{\rm ext}, A_{\rm ext}),$$
the nature of the vacuum is modified by the interactions between the virtual electrons and the external field. In the simplified Furry picture \cite{Furry1}, the vacuum is described through the introduction of the electromagnetic Dirac operator
$$D_{m, e \bA_{\rm ext}} = \bsalpha \cdot (- i \nabla - e A_{\rm ext}) + m \bsbeta + e V_{\rm ext},$$
which is written here in physical units such that the vacuum permittivity $\varepsilon_0$ is equal to $1/4 \pi$. The so-called dressed vacuum is identified to the negative spectral projector
$$P_{m, \bA_{\rm ext}}^{\rm Furry} = \chi_{(- \infty, 0]}(D_{m, e \bA_{\rm ext}}).$$
The virtual electrons of the dressed vacuum span the range of the projector $P_{m, e \bA_{\rm ext}}^{\rm Furry}$. They have no reason to remain identical to the virtual electrons in the free (or bare) vacuum which span the range of the projector $P_{m, 0}^-$. In general, the charge density of the vacuum does not remain constant. The dressed vacuum is polarized. In the case where the external field is strong enough, a positive eigenvalue can even appear in the spectrum of the electromagnetic Dirac operator $D_{m, \bA_{\rm ext}}$, creating an hole in its negative spectrum. A (physical) electron-positron pair is produced, the electron being identified to the positive eigenvalue, while the hole is identified to the positron.

The Furry picture provides a good approximation for the polarization of the vacuum when the external fields are not too strong. In practice, the non-constant charge density of the polarized vacuum modifies the electromagnetic field. The virtual electrons react to the corrected field which in turn affects the nature of the polarization. In order to describe it more accurately, one has to look for more sophisticated models.

\subsection{The Hartree-Fock approximation}
\label{sub:Hartree}

In the Physics literature, vacuum polarization is described using Quantum Electrodynamics which provides extremely accurate computations for the charge and current densities of the polarized vacuum. On the mathematical level, this theory is far from being completely understood, in particular, due to the perturbative nature of its computations. In this subsection, we present a set of approximations which make possible the construction of non-perturbative models for the vacuum polarization. The main one consists in describing the electronic structures as Hartree-Fock states. 

The simplest way to introduce the Hartree-Fock approximation is probably to come back to the description of $N$ classical electrons around a positive density of charge $\nu$. In this situation, the electronic structure is described through the $N$-body Hamiltonian
\begin{equation}
\label{def:H-clas}
H_\nu^N = \sum_{i = 1}^N \bigg( - \frac{1}{2 m} \Delta_{x_i} - e^2 \int_{\R^3} \frac{\nu(y)}{|x_i - y|} dy + \sum_{j > i} \frac{e^2}{|x_i - x_j|} \bigg),
\end{equation}
which acts on the space $L_a^2((\R^3)^N, \C)$ of electronic states $\Psi$ which are anti-symmetric with respect to the permutations of the variables $x_i$ (in order to guarantee the validity of the Pauli exclusion principle), and with a density $|\Psi|^2$ which depends symmetrically on the variables $x_i$ (so as to handle with undistinguishable electrons). The possible electronic structures correspond to the eigenfunctions of the Hamiltonian $H_\nu^N$, the ground state structure corresponding to the minimal eigenvalue. The analysis of $H_\nu^N$ is rather involved (see \cite{HunzSig1, ReedSim4} for more details), so that several approximations have been suggested to simplify the description.

The Hartree-Fock approximation \cite{Hartree1, Fock1} restricts the analysis to the Hartree-Fock states $\Psi$ which write as Slater determinants
$$\Psi(x_1, \ldots, x_N) = \Psi_1 \wedge \cdots \wedge \Psi_N(x_1, \ldots, x_N) = \frac{1}{\sqrt{N !}} \det \big( \Psi_{i}(x_j) \big)_{1 \leq i, j \leq N},$$
where $\Psi_1$, $\ldots$, $\Psi_N$ are $N$ orthogonal wavefunctions in $L^2(\R^3, \C)$. The Hartree-Fock states $\Psi_1 \wedge \cdots \wedge \Psi_N$ are the less correlated electronic structures. Their energy is given by the expression
\begin{align*}
E_\nu^{HF}(\Psi_1 \wedge \cdots \wedge \Psi_N) = & \frac{1}{2 m} \sum_{i = 1}^N \bigg( \int_{\R^3} |\nabla \Psi_i|^2 \bigg) - \frac{e^2}{2} \int_{\R^3} \int_{\R^3} \frac{|\gamma_\Psi(x, y)|^2}{|x - y|} dx dy\\
& - e^2 \int_{\R^3} \int_{\R^3} \frac{\nu(x) \rho_\Psi(y)}{|x - y|} dx dy + \frac{e^2}{2} \int_{\R^3} \int_{\R^3} \frac{\rho_\Psi(x) \rho_\Psi(y)}{|x - y|} dx dy.
\end{align*}
In this formula, $\gamma_\Psi$ refers to the one-body density operator with integral kernel
$$\gamma_\Psi(x, y) = \sum_{i = 1}^N \Psi_i(x) \overline{\Psi_i(y)}.$$
In other words, $\gamma_\Psi$ is the orthogonal projector on the linear space spanned by the wavefunctions $\Psi_i$. Concerning the charge density $\rho_\Psi$, it is equal to
$$\rho_\Psi(x) = \sum_{i = 1}^N |\Psi_i(x)|^2 = \gamma_\Psi(x, x).$$
In particular, the Hartree-Fock energy only depends on the one-body density operator $\gamma_\Psi$ through the identity
\begin{equation}
\label{eq:HF-classic}
\begin{split}
E_\nu^{HF}(\gamma_\Psi) = & \frac{1}{2 m} \tr(- \Delta \, \gamma_\Psi) - \frac{e^2}{2} \int_{\R^3} \int_{\R^3} \frac{|\gamma_\Psi(x, y)|^2}{|x - y|} dx dy\\
& - e^2 \int_{\R^3} \int_{\R^3} \frac{\nu(x) \rho_\Psi(y)}{|x - y|} dx dy + \frac{e^2}{2} \int_{\R^3} \int_{\R^3} \frac{\rho_\Psi(x) \rho_\Psi(y)}{|x - y|} dx dy,
\end{split}
\end{equation}
where $\tr(- \Delta \, \gamma_\Psi)$ is the trace of the finite rank operator $- \Delta \, \gamma_\Psi$. As a consequence, the Hartree-Fock electronic structure can be computed in terms of an orthogonal projector, the Hartree-Fock ground state being identified to the projector which minimizes the energy $E_\nu^{\rm HF}$ among all the possible projectors.

The picture is similar in the relativistic case. In Quantum Electrodynamics, the (formal) Hamiltonian describing an electronic structure in the presence of a (classical) external electromagnetic four-potential $\bA_{\rm ext} = (V_{\rm ext}, A_{\rm ext})$ may be written in the Coulomb gauge as
\begin{equation}
\label{eq:H-QED}
\begin{split}
\H^{\bA_{\rm ext}} = & \int \Psi^*(x) \Big[ \bsalpha \cdot (- i \nabla - e \A(x) - e A_{\rm ext}(x)) + m \bsbeta \Big] \Psi(x) \, dx\\
+ & e \int V_{\rm ext}(x) \rho(x) \, dx + \frac{e^2}{2} \int \int \frac{\rho(x) \rho(y)}{|x - y|} dx \, dy + \frac{1}{8\pi} \int |\curl\A(x)|^2 \, dx.
\end{split}
\end{equation}
In this expression, $\Psi(x)$ refers to the second quantized field operator which annihilates an electron at $x$, while the vector $\A(x)$ is the magnetic field operator for the photons. The density operator $\rho(x)$ is defined as
\begin{equation}
\label{def:rho}
\rho(x) = \frac12\sum_{\sigma = 1}^4 [\Psi_\sigma^*(x), \Psi_\sigma(x)],
\end{equation}
where $[a, b] = a b - b a$ (see \cite{Heisenb2, HeisEul1, Serber2, Schwing1, HaLeSeS1, Lewin1} for more details about the Hamiltonian $\H^{\bA_{\rm ext}}$).

The Hamiltonian $\H^{\bA_{\rm ext}}$ acts on the Fock space $\boF = \boF_{\rm e} \otimes \boF_{\rm ph}$, where $\boF_{\rm e}$ is the fermionic Fock space for the electrons and $\boF_{\rm ph}$ is the bosonic Fock space for the photons. As in the usual case, the main approximation in order to derive Hartree-Fock models for the description of relativistic electrons consists in restricting the Hamiltonian to states of the special form
$$\Omega = \Omega_{\rm HF} \otimes \Omega_{\rm Coh},$$
where $\Omega_{\rm HF}$ is an electronic Hartree-Fock state characterized by its one-body density matrix
$$\gamma(x, y) = \langle \Psi^*(x) \Psi(y) \rangle_{\Omega_{\rm HF}},$$
and $\Omega_{\rm Coh}$ is a coherent state characterized by its classical magnetic potential
$$A(x) = \langle \A(x) \rangle_{\Omega_{\rm Coh}}.$$
In practice, this amounts to representing the electrons by an operator $\gamma$ as in classical Hartree-Fock theory, while the photons are described by a magnetic potential $A$, which is nothing more than a classical vector field. The main difference with the classical case lies in the fact that the operator $\gamma$ does not only take into account the physical electrons, but also the virtual ones representing the vacuum. In general, it is not anymore an orthogonal projector with finite rank, but with infinite one, which causes the apparition of divergences in the computation of the energy. 

Up to a universal constant which diverges in infinite volume, this energy is equal to
\begin{equation}
\label{eq:HF-relativistic}
\begin{split}
\boE_{\rm HF}^{\bA_{\rm ext}}(\gamma, A) = & \tr \Big( D_{m, e (V_{\rm ext}, A + A_{\rm ext})} \big( \gamma - 1/2 \big) \Big) + \frac{e^2}{2} \int_{\R^3} \int_{\R^3} \frac{\rho_{\gamma - 1/2}(x) \rho_{\gamma - 1/2}(y)}{|x - y|} \, dx \, dy\\
& - \frac{e^2}{2} \int_{\R^3} \int_{\R^3} \frac{|(\gamma - 1/2)(x, y)|^2}{|x - y|} \, dx \, dy + \frac{1}{8 \pi} \int_{\R^3}|\curl A(x)|^2 \, dx.
\end{split}
\end{equation}
This expression is similar to the one for the classical Hartree-Fock energy in \eqref{eq:HF-classic}. The kinetic energy is now equal to $\tr(D_{m, 0} (\gamma - 1/2))$, which amounts to replacing the classical operator $- \Delta$ by its relativistic version $D_{m, 0}$. One also recovers the presence of the so-called direct and exchange electrostatic terms in the first, respectively second, line. The main difference lies in the property that the energy is not directly written in terms of the density operator $\gamma$, but in terms of the difference $\gamma - 1/2$. This is a consequence of the charge-conjugation invariant choice \eqref{def:rho} for the density operator $\rho(x)$ (see \cite{HaiLeSo1} for more details). This property is particularly helpful when one attempts to give a rigorous meaning to the expression in \eqref{eq:HF-relativistic}.

The energy in \eqref{eq:HF-relativistic} presents major divergences. Since the operator $D_{m, 0}$ is unbounded and $\gamma - 1/2$ has infinite rank when $\gamma$ is a projector, the kinetic energy is not well-defined. Moreover, since $\gamma - 1/2$ is not a compact operator, defining its kernel $(\gamma - 1/2)(x, y)$, and thereafter the density $\rho_{\gamma - 1/2}$, is a further challenge.

Similar divergences appear in Quantum Electrodynamics, in which regularization techniques have been developed in order to by-pass the difficulty. These techniques mainly rely on the introduction of an ultraviolet cut-off $\Lambda$ in the model. In the sequel, we describe how it is possible to adapt two regularization techniques to provide a rigorous meaning to the energy in \eqref{eq:HF-relativistic} and construct a consistent Hartree-Fock model for the vacuum polarization. This leads to the construction of two different models. The first one is the Bogoliubov-Dirac-Fock model which does not take into account the effects due to the presence of photons, as well as of a possible external magnetic field. In the second one, handling with these effects is in contrast possible due to the use of a more ingenious regularization, the so-called Pauli-Villars regularization.

\section{The Bogoliubov-Dirac-Fock model}
\label{sec:BDF}

\subsection{Derivation and functional framework}
\label{sub:deriv-BDF}

The Bogoliubov-Dirac-Fock model was introduced by Chaix and Iracane in \cite{ChaiIra1} (see also \cite{ChaIrLi1}). It only takes into account kinetic and electrostatic aspects. The external magnetic potential $A_{\rm ext}$ and the magnetic potential $A$ for the photons are set equal to $0$. Concerning the external electrostatic potential $V_{\rm ext}$, it is induced by an external charge density $\nu$ according to the Coulomb formula
$$V_{\rm ext}(x) = - e \int_{\R^3} \frac{\nu(y)}{|x - y|} \, dy.$$
As a result, the formal Hartree-Fock energy for the one-body density operator $\gamma$ reduces to the expression
\begin{equation}
\label{def:E-HF}
\begin{split}
\boE_{\rm HF}^\nu(\gamma) = & \tr \big( D_{m, 0} (\gamma - 1/2) \big) - \alpha \int \int \frac{\rho_{\gamma - 1/2}(x) \nu(y)}{|x - y|} \, dx dy\\
& + \frac{\alpha}{2} \int \int \frac{\rho_{\gamma - 1/2}(x) \rho_{\gamma - 1/2}(y)}{|x - y|} dx dy - \frac{\alpha}{2} \int \int \frac{|(\gamma - 1/2)(x, y)|^2}{|x - y|} \, dx dy,
\end{split}
\end{equation}
where we have introduced the (bare) Sommerfeld fine-structure constant $\alpha = e^2$. Recall that in this formula, the function $(\gamma - 1/2)(x, y)$ refers to the (formal) kernel of the operator $\gamma - 1/2$, while the density $\rho_{\gamma - 1/2}$ is (also formally) defined as
$$\rho_{\gamma - 1/2}(x) = (\gamma - 1/2)(x, x).$$
One can introduce a reduced version of the energy omitting the exchange electrostatic term according to the formula
\begin{equation}
\label{def:E-rHF}
\boE_{\rm rHF}^\nu(\gamma) = \tr \big( D_{m, 0} (\gamma - 1/2) \big) - \alpha \int \int \frac{\rho_{\gamma - 1/2}(x) \nu(y)}{|x - y|} \, dx dy + \frac{\alpha}{2} \int \int \frac{\rho_{\gamma - 1/2}(x) \rho_{\gamma - 1/2}(y)}{|x - y|} \, dx dy.
\end{equation}
In the sequel, we restrict our attention to the reduced energy in order to simplify the analysis. We will point out the results which remain available for the energy with exchange term (see \cite{HaLeSeS1, EstLeSe1, Lewin1} and references therein for more details).

The critical points for the energy $\boE_{\rm rHF}^\nu$ are solutions to the self-consistent equations
$$\big[ \gamma, D_\gamma \big] = 0,$$
where the Fock operator $D_\gamma$ is defined as
$$D_\gamma = D_{m, 0} + \alpha \big( \rho_{\gamma - 1/2} - \nu \big) * \frac{1}{|x|}.$$
An orthogonal projector $\gamma_*$ which minimizes the energy $\boE_{\rm rHF}^\nu$ among all the possible projectors, is solution to the equation
\begin{equation}
\label{eq:rHF}
\gamma_* = \chi_{(- \infty, 0]}(D_{\gamma_*}).
\end{equation}
The projector $\gamma_*$ is interpreted as the polarized vacuum according to the picture of the Dirac sea. The spinors of the virtual electrons in the polarized vacuum span the range of $\gamma_*$. In view of the self-consistent equation \eqref{eq:rHF}, they completely fill in the negative spectrum of the Fock operator $D_{\gamma_*}$.

The model is fashioned to allow for a description of the electronic structure with $N$ physical electrons around the charge density $\nu$. When $\gamma$ is an orthogonal projector with finite rank, its rank, or alternatively its trace, provides the number, or alternatively the total charge, of the electrons in the structure represented by $\gamma$. The ground state structure is defined as a minimizer $\gamma_N$ of the energy $\boE_{\rm rHF}^\nu$ in the charge sector with charge $N$, that is among the orthogonal projectors with trace equal to $N$. Such a minimizer $\gamma_N$ (formally) satisfies the self-consistent equation
\begin{equation}
\label{eq:gamma-N}
\gamma_N = \chi_{(- \infty, \mu_N]}(D_{\gamma_N}),
\end{equation}
where $\mu_N$ is the Fermi level of the electronic structure described by the operator $\gamma_N$. Following again the picture of the Dirac sea, the projector $\gamma_N^{\rm vac} = \chi_{(- \infty, 0]}(D_{\gamma_N})$ describes the virtual electrons of the vacuum polarized by the charge density $\nu$ and the $N$ physical electrons. They are themselves given by the orthogonal projector $\gamma_N^{\rm ph} = \chi_{(0, \mu_N]}(D_{\gamma_N})$. In particular, their energy is positive.

At this stage, it is necessary to emphasize that most of the definitions in the previous discussion are only formal. For example, any orthogonal projectors of finite trace must have a finite rank which cannot be consistent with the identity in \eqref{eq:gamma-N}. However, it is possible to provide a rigorous meaning to the reduced Bogoliubov-Dirac-Fock model (defining for instance a notion of trace for projectors with infinite rank) by invoking regularization techniques.

In this direction, Hainzl, Lewin and Solovej \cite{HaiLeSo1} noticed that the model is well-defined in a finite-dimensional setting. They suggested to restrict the analysis to a box $C_L = [- L/2, L/2[^3$ of size $L$, with periodic boundary conditions, and to introduce an ultraviolet cut-off $\Lambda$ in the Fourier domain. This amounts to assuming that the operators $\gamma$ act on the finite-dimensional space
$$\boH_{L, \Lambda} = \Big\{ \Psi = \sum_{k \in \boR_{L, \Lambda}} a_k \exp i \langle k, \cdot \rangle_{\R^3}, \ a_k \in \C^4 \Big\},$$
where $\boR_{L, \Lambda} = \{ k \in (2 \pi/L) \ \Z^3, \ {\rm s.t.} \ |k| \leq \Lambda \}$. In this periodic framework, the reduced Hartree-Fock energy takes the form
\begin{equation}
\label{def:E-per}
\begin{split}
\boE_{\rm per}^\nu(\gamma) = & \tr \big( D_{m, 0} (\gamma - 1/2) \big) - \alpha \int_{C_L} \int_{C_L} \rho_{\gamma - 1/2}(x) \nu_L(y) G_L(x - y) \, dx dy\\
& + \frac{\alpha}{2} \int_{C_L} \int_{C_L} \rho_{\gamma - 1/2}(x) \rho_{\gamma - 1/2}(y) G_L(x - y) \, dx dy.
\end{split}
\end{equation}
In this definition, the function $\nu_L$ is a periodic version of the charge density $\nu$. It is equal to
$$\nu_L(x) = \bigg( \frac{\sqrt{2 \pi}}{L} \bigg)^3 \sum_{k \in \boR_{L, \Lambda}} \widehat{\nu}(k) \exp i \langle k, x \rangle_{\R^3},$$
where $\widehat{\nu}$ refers to the Fourier transform of $\nu$. In order to get a chance to define properly the Coulomb potentials in \eqref{def:E-per}, it is natural to assume that $\nu$ belongs to the Coulomb space
$$\boC(\R^3) = \Big\{ f \in L^1_{\rm loc}(\R^3, \C), \ {\rm s.t.} \ \| f \|_{\boC}^2 = \int_{\R^3} \frac{|\widehat{f}(k)|^2}{|k|^2} \, dk < + \infty \Big\}.$$
The periodic Coulomb kernel is similarly given by
$$G_L(z) = \frac{4 \pi}{L^3} \underset{k \in (2 \pi/L) \ \Z^3 \setminus \{ 0 \}}{\sum} \bigg( \frac{1}{|k|^2} \exp i \langle k, z \rangle_{\R^3} \bigg) + K_L,$$
where the constant $K_L$ is fixed so that $G_L$ is positive.
Finally, the charge density $\rho_{\gamma - 1/2}$ is defined as
$$\rho_{\gamma - 1/2}(x) = \tr_{\C^4} (\gamma - 1/2)(x, x) = \frac{1}{L^3} \sum_{(j, k) \in (\boR_{L, \Lambda})^2} \tr_{\C^4} \widehat{[\gamma - 1/2]}(j, k) \, \exp i \langle j - k, x \rangle_{\R^3},$$
where $(\gamma - 1/2)(x, y)$ is the kernel of the (finite-rank) operator $\gamma - 1/2$.

In this setting, the periodic reduced Hartree-Fock energy is well-defined on the convex hull of orthogonal projectors on $\boH_{L, \Lambda}$, which is defined as
$$\boG_{L, \Lambda} = \Big\{ \gamma \in \boL(\boH_{L, \Lambda}), \ {\rm s.t.} \ \gamma^* = \gamma \ {\rm and} \ 0 \leq \gamma \leq I \Big\}.$$
Moreover, the energy owns a minimizer on $\boG_{L, \Lambda}$, which can be interpreted as the polarized vacuum
\footnote{Extending the minimization problem under consideration to a convex hull like $\boG_{L, \Lambda}$ is standard in Hartree-Fock theory. The construction of minimizers is simplified, and it turns out that they still solve the initial problem (see e.g. \cite{Lieb2}).}.

\begin{propo}[\cite{HaiLeSo1}]
Let $m$, $\alpha$, $\Lambda$ and $L$ be positive numbers and consider a function $\nu \in \boC(\R^3)$ such that $\widehat{\nu}$ is continuous on $\R^3$. There exists a minimizer $\gamma_L^\nu$ to the minimization problem
$$\inf_{\gamma \in \boG_{L, \Lambda}} \, E_{per}^\nu(\gamma).$$
When $\nu = 0$, the minimizer $\gamma_L^0$ is the negative spectral projector $P_{m, 0, L}^-$ of the restriction of the free Dirac operator $D_{m, 0}$ to the finite-dimensional space $\boH_{L, \Lambda}$.
\end{propo}

The operator $\gamma_L^\nu$ is identified to the polarized vacuum corresponding to the previous periodic setting. A simple way to describe the polarized vacuum in the full space is to consider the thermodynamic limit $L \to + \infty$ of the minimizers $\gamma_L^\nu$. It turns out that the limit exists (when the ultraviolet cut-off $\Lambda$ is fixed) and that it can be described using the reduced Bogoliubov-Dirac-Fock energy (see \cite{HaiLeSo1}).

This energy is defined according to the formula
\begin{equation}
\label{def:E-rBDF}
\boE_{\rm rBDF}^\nu(Q) = \tr_{P_{m, 0}^-} \big( D_{m, 0} Q \big) - \alpha \int_{\R^3} \frac{\nu(x) \rho_Q(y)}{|x - y|} dx dy + \frac{\alpha}{2} \int_{\R^3} \int_{\R^3} \frac{\rho_Q(x) \rho_Q(y)}{|x - y|} dx dy.
\end{equation}
The difference with respect to the reduced Hartree-Fock energy in \eqref{def:E-rHF} lies in the choice of a reference projector to compute the energy. More precisely, the energy is not directly expressed in terms of the projector $\gamma$, but in terms of its difference
$$Q = \gamma - P_{m, 0}^-,$$
with the negative spectral projector $P_{m, 0}^-$ of the free Dirac operator. In other words, the free vacuum energy is set equal to $0$ and all the energies are computed with respect to this reference.

Hainzl, Lewin and S\'er\'e \cite{HaiLeSe1} constructed a functional framework in which the reduced Bogoliu\-bov-Dirac-Fock energy is well-defined and bounded from below. They observed the necessity to conserve an ultraviolet cut-off $\Lambda$ in the model (see \cite{HaiLeSe2}) assuming that the operators $Q$ do not act on $L^2(\R^3, \C^4)$, but on the space
$$\boH_\Lambda = \Big\{ \Psi \in L^2(\R^3, \C^4), \ {\rm s.t.} \ {\rm supp} \big( \widehat{\Psi} \big) \subset B(0, \Lambda) \Big\}.$$
This assumption presents the major advantage to transform the operator $D_{m, 0}$ in a bounded operator. We will see in Subsection \ref{sub:renor} below how to eliminate the ultraviolet cut-off $\Lambda$ from this description of the polarized vacuum.

More precisely, the operators $Q$ belong to the function space
\begin{equation}
\label{def:gS0}
\gS_\Lambda^0 = \Big\{ T \in \gS_2(\boH_\Lambda), \ {\rm s.t.} \ \big( P_{m, 0}^- T P_{m, 0}^-, (I - P_{m, 0}^-) T (I - P_{m, 0}^-) \big) \in \gS_1(\boH_\Lambda)^2 \Big\},
\end{equation}
where $\gS_1(\boH_\Lambda)$ and $\gS_2(\boH_\Lambda)$ are the spaces of trace-class, respectively Hilbert-Schmidt, operators on $\boH_\Lambda$. The trace-class conditions in \eqref{def:gS0} originate in the property that a formal minimizer $Q_*$ of the energy $\boE_{\rm rBDF}^\nu$ is in general not trace class (see Theorem \ref{thm:not-trace-class} below). As a consequence, it is not so easy to define properly its kinetic energy. This is done extending the usual definition through the formula
\begin{equation}
\label{trace}
\tr_{P_{m, 0}^-} \big( T \big) = \tr \big( P_{m, 0}^- T P_{m, 0}^- \big) + \tr \big( (I - P_{m, 0}^-) T (I - P_{m, 0}^-) \big),
\end{equation}
which is well-defined when $T$ is in $\gS_\Lambda^0$. Notice here that the quantity $\tr_{P_{m, 0}^-}(T)$ is equal to the trace of $T$ when $T$ is trace class.

Concerning the charge density $\rho_Q$, its definition relies on the introduction of the ultraviolet cut-off $\Lambda$. Since the operator $Q$ is Hilbert-Schmidt, it owns a kernel $Q(x, y)$, which belongs to the space $L^2(\R^3 \times \R^3)$. Moreover, its Fourier transform is supported into the product set $B(0, \Lambda) \times B(0, \Lambda)$ due to the presence of the cut-off. In particular, it is a smooth function so that the charge density $\rho_Q$ can be defined according to the identity
$$\rho_Q(x) = \tr_{\C^4} \, Q(x, x).$$
With this last definition at hand, one can define properly the reduced Bogoliubov-Dirac-Fock energy
\footnote{The original Bogoliubov-Dirac-Fock energy with exchange term is also well-defined on a set similar to $\gS_\Lambda^0$.}.

\begin{propo}[\cite{HaiLeSe1}]
Let $\Lambda > 0$, $\alpha \geq 0$ and $\nu \in \boC(\R^3)$. The energy $\boE_{rBDF}^\nu$ is well-defined on the set $\gS_\Lambda^0$.
\end{propo}

The introduction of the reduced Bogoliubov-Dirac-Fock model is justified by taking the thermodynamic limit $L \to + \infty$ in the periodic model introduced by Hainzl, Lewin and Solovej in \cite{HaiLeSo1}. As a matter of fact, they established the following theorem (which remains available with a few minor changes in the case of the Bogoliubov-Dirac-Fock model with exchange term). 

\begin{theo}[\cite{HaiLeSo1}]
Let $m$, $\alpha$ and $\Lambda$ be positive numbers and consider a function $\nu \in \boC(\R^3)$ such that $\widehat{\nu}$ is continuous on $\R^3$.\\
$(i)$ When $\nu = 0$, we have
$$\gamma_L^0 = P_{m, 0, L}^-.$$
In particular,
\begin{equation}
\label{eq:conv-0}
\big\| \gamma_L^0 - P_{m, 0}^- \big\|_{\boL(\boH_{L, \Lambda})} \to 0,
\end{equation}
and
\begin{equation}
\label{eq:conv-E-0}
E_{\rm per}^\nu(\gamma_L^0) \to + \infty,
\end{equation}
as $L \to + \infty$.\\
$(ii)$ There exists an operator $Q_*$ in $\gS_\Lambda^0$ such that, up to a possible subsequence, we have
\begin{equation}
\label{eq:conv-nu}
Q_L^\nu(x, y) = \gamma_L^\nu(x, y) - \gamma_L^0(x, y) \to Q_*(x, y),
\end{equation}
as $l \to + \infty$, uniformly on any compact subset of $\R^3 \times \R^3$.\\
$(iii)$ The operator $Q_*$ is a minimizer of the reduced Bogoliubov-Dirac-Fock energy $\boE_{\rm rBDF}^\nu$ on the set
$$\boQ_\Lambda = \Big\{ Q \in \gS_\Lambda^0, \ {\rm s.t.} \ Q^* = Q \ {\rm and} - P_{m, 0}^- \leq Q \leq I - P_{m, 0}^- \Big\}.$$
Moreover, we have
\begin{equation}
\label{eq:conv-E}
\boE_{\rm per}^\nu(\gamma_L^\nu) - \boE_{\rm per}^\nu(\gamma_L^0) \to \boE_{\rm rBDF}^\nu(Q_*) = \min \big\{ E_{\rm rBDF}^\nu(Q), \ Q \in \boQ^\Lambda \big\},
\end{equation}
as $L \to + \infty$.
\end{theo}

The choice of the projector $P_{m, 0}^-$ as reference for a model in the full space is justified by the convergences in \eqref{eq:conv-0} and \eqref{eq:conv-E-0}. In the thermodynamic limit $L \to + \infty$, the free vacuum is given by $P_{m, 0}^-$ according to the original picture of the Dirac sea. Moreover, one has to subtract its infinite energy in order to handle with a reasonable model.

In this case, the convergences in \eqref{eq:conv-nu} and \eqref{eq:conv-E} show that the polarized vacuum is actually described by the reduced Bogoliubov-Dirac-Fock model. More precisely, the polarized vacuum is identified to the projector $\gamma_* = P_{m, 0}^- + Q_*$ corresponding to the minimizer $Q_*$ of the energy $\boE_{\rm rBDF}^\nu$ on the set $\boQ_\Lambda$. Let us recall now some elements about the construction of such a minimizer, as well as its main properties.

\subsection{Construction of the polarized vacuum and the electronic structures with $N$ electrons}
\label{sub:const-BDF}

The existence of the minimizer $Q_*$ was established by Hainzl, Lewin and S\'er\'e in \cite{HaiLeSe1, HaiLeSe2}.

\begin{theo}[\cite{HaiLeSe1, HaiLeSe2}]
Let $\Lambda > 0$, $\alpha \geq 0$ and $\nu \in \boC(\R^3)$.\\
$(i)$ The energy $\boE_{\rm rBDF}^\nu$ is bounded from below on the set $\boQ_\Lambda$. More precisely, we have
\begin{equation}
\label{eq:Bach}
\boE_{\rm rBDF}^\nu(Q) + \frac{\alpha}{2} \int_{\R^3} \int_{\R^3} \frac{\nu(x) \nu(y)}{|x - y|} \, dx dy \geq 0,
\end{equation}
for any $Q \in \boQ_\Lambda$. When $\nu = 0$, inequality \eqref{eq:Bach} turns into an equality if and only if $Q = 0$.\\
$(ii)$ The energy $\boE_{\rm rBDF}^\nu$ owns a minimizer $Q_*$ on $\boQ_\Lambda$, which is solution to the self-consistent equation
\begin{equation}
\label{eq:Q_*}
\gamma_* = Q_* + P_{m, 0}^- = \chi_{(- \infty, 0)} \big( D_* \big) + \delta_*,
\end{equation}
where the Fock operator $D_*$ is equal to
$$D_* = D_{m, 0} + \alpha \big( \rho_{Q_*} - \nu \big) * \frac{1}{|x|},$$
while $\delta_*$ refers to a finite-rank operator with range in the kernel of $D_*$.\\
$(iii)$ The charge density $\rho_{Q_*}$ is uniquely determined. When the external charge density $\nu$ satisfies the condition
\begin{equation}
\label{eq:cond-unique}
2^\frac{11}{6} \pi^\frac{1}{6} \alpha \| \nu \|_{\boC} < m^\frac{1}{2},
\end{equation}
the minimizer $Q_*$ is also unique. In this case, the operator $\delta_*$ is equal to $0$, and
\begin{equation}
\label{eq:neutre}
N_* = \tr_{P_{m, 0}^-} \big( Q_* \big) = 0.
\end{equation}
\end{theo}

\begin{rem}
The previous theorem extends with a few minor changes to the Bogoliubov-Dirac-Fock model with exchange term (see \cite{HaiLeSe1, HaiLeSo1}). In this case, inequality \eqref{eq:Bach} was first proved in \cite{BaBaHeS1}.
\end{rem}

Once again, the fact that $Q = 0$ is the unique minimizer of the energy $\boE_{\rm rBDF}^0$ is consistent with the picture of the Dirac sea for the free vacuum. In presence of an external charge density $\nu$, the polarized vacuum is described by the operator $\gamma_* = Q_* + P_{m, 0}^-$, which is not necessarily an orthogonal projector due to the presence of the finite-rank operator $\delta_*$. This defect is a common drawback of the reduced Hartree-Fock models (see e.g. \cite{Solovej1}). It is possible to withdraw the operator $\delta_*$ from the self-consistent equation satisfied by a minimizer of the Bogoliubov-Dirac-Fock energy with exchange term (see \cite{HaiLeSe1, HaiLeSo1}).

Under condition \eqref{eq:cond-unique}, the kernel of the Fock operator $D_*$ is necessarily equal to $0$, so that the operator $\delta_*$ identically vanishes. In this case, one recovers the picture of the Dirac sea in which the polarized vacuum is identified to the negative spectral projector $\gamma_*$ of the Fock operator $D_*$. It follows from \eqref{eq:neutre} that the polarized vacuum is globally neutral. The trace of $Q_*$ is indeed interpreted as the charge of the electronic structure represented by $\gamma_*$ since, at least formally, the charge of the negative projector $P_{m, 0}^-$, which gives account of the free vacuum, must be equal to $0$. This is consistent with the observation that a weak electrostatic potential cannot produce physical electrons in the vacuum.

In another direction, one can ask for the behaviour of the minimizer $Q_*$ when $\Lambda \to + \infty$. It turns out that the model collapses in this limit.

\begin{propo}[\cite{HaiLeSe2}]
Let $\alpha \geq 0$ and $\nu \in \boC(\R^3)$. We denote by $Q_*^\Lambda$ a minimizer of the energy $\boE_{\rm rBDF}^\nu$ on $\boQ_\Lambda$ for a positive number $\Lambda$. Then,
$$\big\| |D_{m, 0}|^\frac{1}{2} Q_*^\Lambda \big\|_{\gS_2} \to 0 \quad {\rm and} \quad \alpha \big\| \rho_{Q_*^\Lambda} - \nu \big\|_{\boC} \to 0,$$
as $\Lambda \to + \infty$. In particular,
\begin{equation}
\label{Enul}
\boE_{\rm rBDF}^\nu(Q_*^\Lambda) \to - \frac{\alpha}{2} \int_{\R^3} \int_{\R^3} \frac{\nu(x) \nu(y)}{|x - y|} dx dy,
\end{equation}
as $\Lambda \to + \infty$.
\end{propo}

In the limit $\Lambda \to + \infty$, the operators $Q_*^\Lambda$ converge to $0$, while their charge densities $\rho_{Q_*^\Lambda}$ tend to the function $\nu$. The limit density is not the charge density of the limit operator. In other words, the model does not remain physically consistent in the limit $\Lambda \to + \infty$. In particular, taking this limit is not a way to eliminate the ultraviolet cut-off $\Lambda$. This property is related to the Landau pole phenomenon which was originally described by Landau \cite{Landau1}, and Landau and Pomeranchuk \cite{LandPom1}. In Subsection \ref{sub:renor}, we will see how to deal with the ultraviolet cut-off by invoking renormalization arguments from Quantum Electrodynamics.

Concerning the description of the electronic structure with $N$ electrons, recall that they are described by the minimizers $Q_N$ of the reduced Bogoliubov-Dirac-Fock energy $\boE_{rBDF}^\nu$ in the charge sectors
$$\boQ_\Lambda(N) = \big\{ Q \in \boQ_\Lambda, \ {\rm s.t.} \ \tr_{P_{m, 0}^-}(Q) = N \big\}.$$
Solving this minimization problem is more involved than the construction of the polarized vacuum. The main difficulty
arises in the fact that the charge sectors $\boQ_\Lambda(N)$ are not stable under weak convergence in $\gS_\Lambda^0$. To our knowledge, the construction of the electronic structure with $N$ electrons remains open for the Bogoliubov-Dirac-Fock model with exchange term. However, Hainzl, Lewin and S\'er\'e \cite{HaiLeSe3} proved the following Hunziker-van Winter-Zhislin condition (for both the reduced and original models).

\begin{propo}[\cite{HaiLeSe3}]
Let $\Lambda > 0$, $\alpha \geq 0$ and $\nu \in \boC(\R^3)$, and set
\begin{equation}
\label{def:inf-N}
E_{\rm rBDF}^\nu(q) = \inf_{Q \in \boQ_\Lambda(q)} \boE_{\rm rBDF}^\nu(Q),
\end{equation}
for any real number $q$.\\
$(i)$ Let $q \in \R$. If
\begin{equation}
\label{eq:HVZ}
E_{\rm rBDF}^\nu(q) < E_{\rm rBDF}^\nu(q - k) + E_{\rm rBDF}^0(k),
\end{equation}
for any $k \in \R^*$, then the minimization problem \eqref{def:inf-N} owns a minimizer $Q_q$.\\
$(ii)$ In case of existence, a minimizer $Q_q$ is solution to the self-consistent equation
\begin{equation}
\label{eq:Q_q}
\gamma_q = Q_q + P_{m, 0}^- = \chi_{(- \infty, \mu_q)} \big( D_q \big) + \delta_q,
\end{equation}
where $\mu_q \in [- 1, 1]$ is the Fermi level of the electronic structure. The Fock operator $D_q$ is defined as
$$D_q = D_{m, 0} + \alpha \big( \rho_{Q_q} - \nu \big)*\frac{1}{|x|}.$$
The self-adjoint operator $\delta_q$ is finite rank when $\mu_q \neq \pm 1$, trace class otherwise, and it ranges in the kernel of the operator $D_q - \mu_q I$.
\end{propo}

When $q = N$ is a positive integer, the polarized vacuum is interpreted as the spectral projector $\gamma_N^{\rm vac} = \chi_{(- \infty, 0]}(D_N)$, while the electronic structure with $N$ physical electrons is described by the spectral projector $\gamma_N^{\rm ph} = \chi_{(0, \mu_N]}(D_N)$. When the electrostatic potential induced by the charge density $\nu$ is weak enough, that is under condition \eqref{eq:cond-unique}, the equality
$$\tr_{P_{m, 0}^-} \big( \gamma_N^{\rm vac} - P_{m, 0}^- \big) = 0,$$
holds, which means that the polarized vacuum is neutral. In this case,
$$\tr_{P_{m, 0}^-} \big( \gamma_N^{\rm ph} \big) = \tr_{P_{m, 0}^-} \big( \gamma_N - P_{m, 0}^- \big) = \tr_{P_{m, 0}^-} \big( Q_N \big) = N.$$
In other words, the electronic structure really exhibits $N$ physical electrons which fill in positive energy levels of the Fock operator $D_N$.

In contrast, nothing rules out the possibility that $q$ is a negative integer. In this situation, the minimizer $Q_q$ describes a positronic structure with $|q|$ positrons. The Fermi level $\mu_q$ is negative. The polarized vacuum is identified as before to the spectral projector $\gamma_q^{\rm vac} = \chi_{(- \infty, 0]}(D_q)$, while the $|q|$ positrons are represented by the spectral projector $\gamma_q^{\rm ph} = \chi_{(\mu_q, 0]}(D_q)$.

Checking the validity of condition \eqref{eq:HVZ} for the Bogoliubov-Dirac-Fock model with exchange term is widely open except for some weak-coupling or non-relativistic regimes (see \cite{HaiLeSe3} for more details). As a consequence, the existence for a given value of $N$ of electronic structures with $N$ electrons also remains an open problem. For the reduced model, it is possible to characterize precisely the numbers for which inequality \eqref{eq:HVZ} is fulfilled.

\begin{thm}[\cite{GraLeSe1}]
\label{thm:mini-rBDF}
Let $\Lambda > 0$, $\alpha \geq 0$ and $\nu \in L^1(\R^3) \cap \boC(\R^3)$. Set
$$Z = \int_{\R^3} \nu.$$
$(i)$ There exist two constants $q_m \in [- \infty, + \infty]$ and $q_M \in [q_m, + \infty]$ such that the minimization problem \eqref{def:inf-N} owns a minimizer $Q_q$ if and only if
$$q_m \leq q \leq q_M.$$
$(ii)$ Let
$$q_* = \tr_{P_{m, 0}^-}(Q_*),$$
where $Q_*$ is the absolute minimizer of the energy $\boE_{\rm rBDF}^\nu$ on $\boQ_\Lambda$. Then,
$$(q_*, Z) \in [q_m, q_M]^2.$$
In particular, when condition \eqref{eq:cond-unique} is fulfilled, the minimization problem \eqref{def:inf-N} owns a minimizer $Q_q$ for any value of $q \in [0, Z]$.
\end{thm}

When the electrostatic potential induced by the charge density $\nu$ is weak enough, Theorem \ref{thm:mini-rBDF} guarantees the existence of electronic structures with $N$ electrons for any integer $N$ between $0$ and $Z$. This is physically relevant in the sense that structures with $N$ electrons are experimentally observed for $N$ between $0$ and $Z + 1$. On the other hand, Theorem \ref{thm:mini-rBDF} does not rule out the existence of electronic structures with an arbitrary number of electrons. To our knowledge, there are indeed no available upper bounds on $|q_m|$ and $q_M$.

Computing such ionization bounds is quite involved even for classical Hartree-Fock models (see \cite{Solovej1, Solovej2}). Concerning the reduced Bogoliubov-Dirac-Fock model, an additional difficulty lies in the sharpness of the ultraviolet cut-off $\Lambda$. Imposing such a sharp cut-off amounts to replacing the free Dirac operator $D_{m, 0}$ by the operator $D_{m, 0}^\Lambda$ with Fourier transform
\begin{equation}
\label{eq:D-Lambda}
\widehat{D_{m, 0}^\Lambda}(p) = \big( \bsalpha \cdot p + m \beta \big) \big( 1 + \chi(|p|^2/\Lambda^2) \big),
\end{equation}
where
$$\chi(r) = 0 \ {\rm if} \ 0 \leq r < 1, \quad {\rm and} \quad \chi(r) = + \infty \ {\rm if} \ r \geq 1.$$
The discontinuity in the Fourier transform of $D_{m, 0}^\Lambda$ is one of the source of troubles which prevents from computing upper bounds on $|q_m|$ and $q_M$.

A natural way to by-pass the difficulty consists in replacing the function $\chi$ in \eqref{eq:D-Lambda} by a smooth function. This has no major consequences on the previous analysis of the reduced Bogoliubov-Dirac-Fock model (see \cite{GraLeSe1} for more details). In particular, Theorem \ref{thm:mini-rBDF} remains available with a smooth ultraviolet cut-off. Moreover, for the special choice
$$\chi(r) = r^2,$$
and for $\alpha$, $\alpha \| \nu \|_\boC$ and $\alpha \ln \Lambda$ small enough, it is possible to compute the bounds
\begin{equation}
\label{eq:ionis1}
- K_{\alpha, \nu, \Lambda} \leq q_m \leq 0,
\end{equation}
and
\begin{equation}
\label{eq:ionis2}
Z \leq q_M \leq 2 Z + K_{\alpha, \nu, \Lambda}.
\end{equation}
At the non-relativistic limit $\alpha \to 0$, $\Lambda \to + \infty$ and $\alpha \ln \Lambda \to 0$, the positive number $K_{\alpha, \nu, \Lambda}$ satisfies
$$K_{\alpha, \nu, \Lambda} \to 0.$$
In this limit, one recovers the bound computed by Lieb \cite{Lieb1} in the classical case, that is
$$q_m = 0 \quad {\rm and} \quad Z \leq q_M \leq 2 Z.$$
We refer to \cite{GraLeSe1} for more precise statements about this topic, as well as for the proofs of \eqref{eq:ionis1} and \eqref{eq:ionis2} (which are essentially based on the arguments developed by Lieb in \cite{Lieb1}).

\subsection{Charge renormalization for the polarized vacuum}
\label{sub:renor}

A crucial ingredient in the proof of Theorem \ref{thm:mini-rBDF} is the following proposition.

\begin{thm}[\cite{GraLeSe1}]
\label{thm:not-trace-class}
Let $\Lambda > 0$ and $\alpha \geq 0$. Consider a function $\nu \in L^1(\R^3) \cap \boC(\R^3)$, with
$$\int_{\R^3} \nu = Z \in \R,$$
and denote by $q_m$ and $q_M$, the two extremal values in Theorem \ref{thm:mini-rBDF} for the interval of existence of a minimizer $Q_q$ of the energy $E_{\rm rBDF}^\nu$ on the charge sector $\boQ_\Lambda(q)$. Given any number $q \in [q_m, q_M]$, the charge density $\rho_{Q_q}$ is an integrable function on $\R^3$. Its integral is given by
\begin{equation}
\label{int:rho-q}
\int_{\R^3} \big( \nu - \rho_{Q_q} \big) = \frac{Z - q}{1 + \alpha B_\Lambda^0},
\end{equation}
where
\begin{equation}
\label{def:B-Lambda-0}
B_\Lambda^0 = \frac{1}{\pi} \int_0^\frac{\Lambda}{\sqrt{1 + \Lambda^2}} \frac{z^2 - \frac{z^4}{3}}{1 - z^2} \, dz.
\end{equation}
\end{thm}

In view of Theorem \ref{thm:not-trace-class}, a minimizer $Q_q$ is not a trace-class operator, at least when $q \neq Z$. Otherwise, the integral of its density would be equal to
$$\int_{\R^3} \rho_{Q_q} = \tr Q_q = q,$$
which contradicts the fact that the number $B_\Lambda^0$ in \eqref{def:B-Lambda-0} is positive. 

The fact that the minimizers $Q_q$ are not trace class generates a difficulty in their physical interpretation. The total electrostatic potential $V_{\rm ph}$ which is induced by the external charge density $\nu$ and the electrons represented by $Q_q$, is defined by the Coulomb formula as
\begin{equation}
\label{def:V-ph}
V_{\rm ph}(x) = \alpha \int_{\R^3} \frac{\nu(y) - \rho_{Q_q}(y)}{|x - y|} \, dy.
\end{equation}
When $|x|$ is large enough, an approximation for the potential $V_{\rm ph}$ is provided by the expression
$$V_{\rm ph}(x) \approx \frac{\alpha}{|x|} \int_{\R^3} \big( \nu - \rho_{Q_q} \big).$$
In view of \eqref{int:rho-q}, it follows that
\begin{equation}
\label{eq:V-ph-infty}
V_{\rm ph}(x) \approx \frac{\alpha}{1 + \alpha B_\Lambda^0} \frac{Z - q}{|x|} \quad \neq \quad \alpha \frac{Z - q}{|x|},
\end{equation}
when $|x| \to + \infty$. Whereas the minimizer $Q_q$ is supposed to represent an electronic structure with $q$ electrons, the potential $V_{\rm ph}$ does not match the Coulomb formula for a potential induced by a total charge equal to $Z - q$.

At this stage, one could argue that it is sufficient to take the limit $\Lambda \to + \infty$ to solve the difficulty. This is not the case. The constant $B_\Lambda^0$ is logarithmically divergent when $\Lambda \to + \infty$. One can check that
\begin{equation}
\label{eq:dev-B-Lambda}
B_\Lambda^0 = \frac{2}{3 \pi} \ln(\Lambda) - \frac{5}{9 \pi} + \frac{2}{3 \pi} \ln(2) + \underset{\Lambda \to + \infty}{\boO} \Big( \frac{1}{\Lambda^2} \Big).
\end{equation}
As a consequence, the potential $V_{\rm ph}$ vanishes in the limit $\Lambda \to + \infty$. This is another sign of the collapse of the model in this limit.

This logarithmic divergence also appears in Quantum Electrodynamics. This difficulty is solved by introducing a charge renormalization. Roughly speaking, charge renormalization consists in accepting the idea that the bare fine-structure constant $\alpha$ in the model is not the fine-structure constant $\alpha_{\rm ph}$ which is experimentally observed. The physical fine-structure constant is defined according to the formula
\begin{equation}
\label{def:alpha-ph}
\alpha_{\rm ph} = \frac{\alpha}{1 + \alpha B_\Lambda^0},
\end{equation}
so that formula \eqref{eq:V-ph-infty} matches with the limit at infinity of a Coulomb potential induced by a total charge equal to $Z - q$.

Modifying the definition of the fine-structure constant affects in turn the definition of the potential $V_{\rm ph}$ in the sense that this potential must be equal to
\begin{equation}
\label{eq:V-ph}
V_{\rm ph}(x) = \alpha \int_{\R^3} \frac{\rho_{\rm ph}(y)}{|x - y|} \, dy,
\end{equation}
where $\rho_{\rm ph}$ refers to the total charge density which is experimentally observed. In view of \eqref{def:V-ph}, one has to admit that the value of $\rho_{\rm ph}$ is equal to
\begin{equation}
\label{def:rho-ph}
\alpha_{\rm ph} \rho_{\rm ph} = \alpha_{\rm ph} \big( \nu - \rho_{Q_q} \big).
\end{equation}

A natural question is to ask for the physical relevance of the quantities $\alpha_{\rm ph}$ and $\rho_{\rm ph}$. Using Quantum Electrodynamics, one can compute a power series of $\rho_{\rm ph}$ with respect to $\alpha_{\rm ph}$ (tending to $0$), and check that the resulting computations are in agreement with physical measurements. In the case of the reduced Bogoliubov-Dirac-Fock model, one can ask for a similar property: does it remain possible to compute an expansion of $\rho_{\rm ph}$ when $\alpha_{\rm ph} \to 0$, and to verify the consistence of the expansion with respect to the one provided by Quantum Electrodynamics ? The answer is positive provided one introduces a multiplicative renormalization as in Quantum Electrodynamics. 

As a matter of fact, our model still contains an ultraviolet cut-off $\Lambda$ whose value is unknown. In Quantum Electrodynamics, multiplicative renormalization consists in fixing the value of $\Lambda$ so that the replacement of the bare fine-structure constant $\alpha$ by the physical one $\alpha_{\rm ph}$ amounts to a change of physical units (see \cite{Dyson0} for more details). In practice, the value of $\alpha_{\rm ph}$ is set equal to
\begin{equation}
\label{def:Z3}
\alpha_{\rm ph} = Z_3 \alpha,
\end{equation}
where $Z_3$ is a fixed positive number. The computations of the power series of $\rho_{\rm ph}$ with respect to $\alpha_{\rm ph}$ are made for $Z_3$ fixed. This amounts to considering $\Lambda$ as a function of $\alpha_{\rm ph}$ and $Z_3$ according to the identity
\begin{equation}
\label{eq:Z3}
\alpha_{\rm ph} B_\Lambda^0 = 1 - Z_3.
\end{equation}
The coefficients of the resulting power series are surprisingly independent of the value of $Z_3$. In other words, the perturbative computation of the density $\rho_{\rm ph}$ is independent of the ultraviolet cut-off $\Lambda$ provided it is fixed according to \eqref{eq:Z3}. Describing perturbatively the polarized vacuum or an electronic structure with $N$ electrons does not really require to set the value of $\Lambda$.

This property remains true for the polarized vacuum when it is described by the reduced Bogoliubov-Dirac-Fock model.

\begin{thm}[\cite{GraLeSe2}]
\label{thm:renor}
Let $m \in \N$ and $\nu \in L^2(\R^3) \cap \boC(\R^3)$ such that
$$\int_{\R^3} \ln \big( 1 + |k| \big)^{2 m + 2} |\widehat{\nu}(k)|^2 \, dk < + \infty.$$
We denote by $\rho_{\rm ph}(\alpha_{\rm ph}, Z_3)$, the physical density associated to the minimizer $Q_*$ of the energy $E_{\rm rBDF}^\nu$ on $\boQ_\Lambda$ according to formulae \eqref{def:alpha-ph}, \eqref{def:rho-ph} and \eqref{def:Z3}\\
$(i)$ Let $0 < \epsilon < 1/2$. There exist two positive numbers $K$ and $a$, depending only on $m$, $\epsilon$ and $\nu$, and a family of functions $(\nu_n)_{0 \leq n \leq m}$ in $L^2(\R^3) \cap \boC(\R^3)$, such that
\begin{equation}
\label{eq:asymptotics}
\Big\| \rho_{\rm ph}(\alpha_{\rm ph}, Z_3) - \sum_{n = 0}^m \nu_n \alpha_{\rm ph}^n \Big\|_{L^2 \cap \boC} \leq K \alpha_{\rm ph}^{m + 1},
\end{equation}
for any $0 \leq \alpha_{\rm ph} \leq a$ and any $\epsilon \leq Z_3 \leq 1 - \epsilon$.\\
$(ii)$ The function $\nu_0$ is identically equal to $\nu$, while the functions $\nu_n$ are inductively defined as
\begin{equation}
\label{def:nu-n}
\nu_1 = \boU(\nu_0), \quad {\rm and} \quad \nu_n = \boU(\nu_{n - 1}) + \underset{j = 3}{\overset{n}{\sum}} \ \underset{n_1 + \cdots + n_j = n - j}{\sum} F_j \big( \nu_{n_1}, \ldots, \nu_{n_j} \big),
\end{equation}
for $n \geq 2$. In this expression, $\boU$ refers to the Uehling operator defined as the Fourier multiplier corresponding to the function
\begin{equation}
\label{def:Uehling}
U(k) = \frac{|k|^2}{4 \pi} \int_0^1 \frac{z^2 - \frac{z^4}{3}}{1 + \frac{|k|^2 (1 - z^2)}{4}} \, dz = \frac{12 - 5 |k|^2}{9 \pi |k|^2} + \frac{\sqrt{4 + |k|^2}}{3 \pi |k|^3} \big( |k|^2 - 2 \big) \ln \Big( \frac{\sqrt{4 + |k|^2} + |k|}{\sqrt{4 + |k|^2} - |k|} \Big).
\end{equation}
The nonlinear maps $F_j(\mu_1, \ldots, \mu_j)$ are equal to the charge densities of the operators
$$Q_j(\mu_1, \ldots, \mu_j) = \frac{1}{2 \pi} \int_{- \infty}^{+ \infty} \frac{1}{D_{m, 0} + i \eta} \prod_{n = 1}^j \Big( \mu_n * \frac{1}{|\cdot|} \ \frac{1}{D_{m, 0} + i \eta} \Big) \, d\eta.$$
$(iii)$ In particular, the functions $\nu_n$ do not depend on $Z_3$, but only on the external charge density $\nu$.
\end{thm}

Beyond the fact that they do not depend on $Z_3$, the values of the densities $\nu_n$ are consistent with the perturbative computations of Quantum Electrodynamics. The function $\nu_0$ is equal to the external charge density $\nu$. This is exactly the total charge density of the system in the non-relativistic case. The function $\nu_1$ induces a Coulomb potential equal to the Uehling potential (see \cite{Serber1, Uehling1}) given by
$$V_{\rm Ueh}(x) = \alpha_{\rm ph}^2 \nu_1 * \frac{1}{|x|} = \frac{\alpha_{\rm ph}^2}{3 \pi} \int_1^{+ \infty} (t^2 - 1)^\frac{1}{2} \Big( \frac{2}{t^2} + \frac{1}{t^4} \Big) \bigg( \int_{\R^3} e^{- 2 |x - y| t} \frac{\nu(y)}{|x - y|} \, dy \bigg) \, dt.$$
This potential is the first correction of the polarized vacuum density which is computed by Quantum Electrodynamics.

The proof of Theorem \ref{thm:renor} is based on equation \eqref{eq:Q_*}. Assuming that condition \eqref{eq:cond-unique} is satisfied, the operator $\delta_*$ identically vanishes, so that we can invoke the Cauchy formula to write
\begin{equation}
\label{eq:Q*}
Q_* = - \frac{1}{2 \pi} \int_{- \infty}^{+ \infty} \Big( \frac{1}{D_* + i \eta} - \frac{1}{D_{m, 0} + i \eta} \Big) d\eta = \sum_{j = 1}^{+ \infty} \alpha^j Q_{\Lambda, j}.
\end{equation}
Here, the operators $Q_{\Lambda, j}$ are given by
$$Q_{\Lambda, j} = - \frac{1}{2 \pi} \int_{- \infty}^{+ \infty} \frac{1}{D_{m, 0} + i \eta} \Big( \Pi_\Lambda \, \big( \nu - \rho_{Q_*} \big) * \frac{1}{|\cdot|} \, \Pi_\Lambda \, \frac{1}{D_{m, 0} + i \eta} \Big)^k \, d\eta.$$
Their dependence with respect to the ultraviolet cut-off $\Lambda$ is explicit through the truncation operators $\Pi_\Lambda$ defined as
$$\widehat{\Pi_\Lambda(f)} = \widehat{f} \, 1_{B(0, \Lambda)},$$
for any $f \in L^2(\R^3, \C^4)$.
 
Translated in terms of the Fourier transforms of the densities $\rho_{Q_*}$ and $\rho_{Q_{\Lambda, j}}$, expansion \eqref{eq:Q*} writes as
\begin{equation}
\label{eq:rho-Q*}
\widehat{\rho_{Q_*}}(k) = - \alpha B_\Lambda(k) \big( \widehat{\rho_{Q_*}}(k) - \widehat{\nu}(k) \big) + \widehat{F_\Lambda} \big( \alpha (\nu - \rho_{Q_*}) \big)(k).
\end{equation}
In this expression, the Fourier multiplier $B_\Lambda$ is equal to
$$B_\Lambda(k) = \frac{1}{\pi} \int_0^{Z_\Lambda(|k|)} \frac{z^2 - \frac{z^4}{3}}{(1 - z^2) \big( 1 + \frac{|k|^2}{4} (1 - z^2) \big)} \, dz + \frac{|k|}{2 \pi} \int_0^{Z_\Lambda(|k|)} \frac{z - \frac{z^3}{3}}{\sqrt{1 + \Lambda^2} - \frac{|k|}{2} z} \, dz,$$
where we have set
$$Z_\Lambda(r) = \frac{\sqrt{1 + \Lambda^2} - \sqrt{1 + (\Lambda - r)^2}}{r}.$$
It is useful to write the function $B_\Lambda$ as
$$B_\Lambda(k) = B_\Lambda^0 - U_\Lambda(k),$$
where $B_\Lambda^0 = B_\Lambda(0)$ is defined in \eqref{def:B-Lambda-0} (see \cite{PaulRos1}). 

The nonlinear map $F_\Lambda$ in \eqref{eq:Q*} is defined as
\begin{equation}
\label{def:F-Lambda}
F_\Lambda(\mu) = \sum_{n \geq 1} F_{\Lambda, 2 n + 1}(\mu, \ldots, \mu),
\end{equation}
where $F_{\Lambda, n}(\mu_1, \ldots, \mu_n)$ stands for the charge density of the operator
$$T_{\Lambda, n}(\mu_1, \ldots, \mu_n) = \frac{1}{2 \pi} \int_{- \infty}^{+ \infty} \frac{1}{D_{m, 0} + i \eta} \prod_{j = 1}^n \Big( \Pi_\Lambda \, \mu_j * \frac{1}{|\cdot|} \, \Pi_\Lambda \, \frac{1}{D_{m, 0} + i \eta} \Big) \, d\eta.$$
In particular, the functions $F_{\Lambda, n}$ identically vanish when $n$ is even (see \cite{Furry1}).

In the physical variables $\alpha_{\rm ph}$ and $\rho_{\rm ph}$, equation \eqref{eq:Q*} reduces to
\begin{equation}
\label{eq:rho-ph}
\big( 1 - \alpha_{\rm ph} U_\Lambda(k) \big) \widehat{\rho_{\rm ph}}(k) + \widehat{F}_\Lambda \big( \alpha_{\rm ph} \rho_{\rm ph} \big) (k) = \widehat{\nu_\Lambda}(k),
\end{equation}
with $\widehat{\nu_\Lambda}(k) = \widehat{\nu}(k) \, 1_{B(0, 2 \Lambda)}(k)$. At this stage, it is possible to substitute in \eqref{eq:rho-ph} the function $\rho_{\rm ph}$ by the formal expansion
\begin{equation}
\label{eq:dev-rho-ph}
\rho_{\rm ph} = \sum_{n \geq 0} \nu_{\Lambda, n} \alpha_{\rm ph}^n,
\end{equation}
and compute the value of the coefficients $\nu_{\Lambda, n}$. They are inductively given by
\begin{equation}
\label{def:nu-nLambda}
\nu_{\Lambda,0} = \nu_\Lambda, \ \nu_{\Lambda, 1} = \boU_\Lambda(\nu_\Lambda), \quad {\rm and} \quad \nu_{\Lambda, n} = \boU_\Lambda(\nu_{\Lambda, n - 1}) + \underset{j = 3}{\overset{n}{\sum}} \ \underset{n_1 + \cdots + n_j = n - j}{\sum} F_{\Lambda, j} \big( \nu_{\Lambda, n_1}, \ldots, \nu_{\Lambda, n_j} \big),
\end{equation}
for any $n \geq 2$. In this expression, $\boU_\Lambda$ refers to the Fourier multiplier associated to the function $U_\Lambda$.

Expansion \eqref{eq:asymptotics} is then proved into two steps. The formal expansion in \eqref{eq:dev-rho-ph} is first rigorously derived as a Taylor expansion of order $m$. This is summarized by the inequality
\begin{equation}
\label{eq:dev-rho-Lambda}
\Big\| \rho_{\rm ph} - \sum_{n = 0}^m \nu_{\Lambda, n} \alpha_{\rm ph}^n \Big\|_{L^2 \cap \boC} \leq K \alpha_{\rm ph}^{m + 1},
\end{equation}
for $\alpha_{\rm ph}$ small enough. On the other hand, one can check that the coefficients $\nu_{\Lambda, n}$ converge at the limit $\Lambda \to + \infty$ to the functions $\nu_n$ defined in \eqref{def:nu-n}. This follows from the convergences of the functions $U_\Lambda$ and $F_{\Lambda, j}$ to $U$, respectively $F_j$, in the limit $\Lambda \to + \infty$. In particular, one can replace the coefficients $\nu_{\Lambda, n}$ in \eqref{eq:dev-rho-Lambda} by the functions $\nu_n$, so as to obtain expansion \eqref{eq:asymptotics}.

However, we are not interested in the limit $\Lambda \to + \infty$, but in the limit $\alpha_{\rm ph} \to 0$, with an ultraviolet cut-off $\Lambda$ fixed so that $Z_3 = 1 - \alpha_{\rm ph} B_\Lambda$ remains constant. This last assumption is the crucial ingredient in the proof. In view of \eqref{eq:dev-B-Lambda}, this amounts to assuming that
\begin{equation}
\label{eq:dev-Lambda}
\Lambda \approx \exp \frac{3 \pi Z_3}{2 \alpha_{\rm ph}},
\end{equation}
when $\alpha_{\rm ph} \to 0$. As a consequence, the Taylor series with respect to $\alpha_{\rm ph}$ of any negative powers of $\Lambda$ identically vanish. In other words, terms controlled by inverse powers of $\Lambda$ play no role in the expansion of $\rho_{\rm ph}$ with respect to $\alpha_{\rm ph}$. In particular, one can check that
$$\| \nu_{\Lambda, n} - \nu_n \|_{L^2 \cap \boC} \leq K \alpha_{\rm ph}^{m + 1 - n},$$
for any $0 \leq n \leq m$. Expansion \eqref{eq:asymptotics} follows combining with \eqref{eq:dev-rho-Lambda}.

To conclude the derivation of Theorem \ref{thm:renor}, notice that the dependence on $Z_3$ is entirely contained in \eqref{eq:dev-Lambda}, so that it vanishes when a Taylor expansion of an inverse power of $\Lambda$ is performed. This explains why the coefficients $\nu_n$ do not depend on $Z_3$.

\section{The Pauli-Villars regulated model}
\label{sec:PV}

\subsection{Formal derivation}
\label{sub:deriv-PV}

A major restriction in the (reduced) Bogoliubov-Dirac-Fock model lies in its purely electrostatic nature which prevents, for instance, from describing the role played by photons in the vacuum polarization. Our goal is now to explain, at least formally, how to derive a more general Hartree-Fock model taking into account some features of the photons as well as of external magnetic fields. We refer to \cite{GrHaLeS1} for more details (see also \cite{Lewin2}). 

The main difficulty arises in the choice of the regularization which we have to introduce in order to define the model properly. Recall that the formal Hartree-Fock energy for describing the polarized vacuum in the Coulomb gauge may be written according to \eqref{eq:HF-relativistic} as
\begin{align*}
\boE_{HF}^{\bA_{\rm ext}}(\gamma, A) = & \tr \Big( D_{m, e (V_{\rm ext}, A + A_{\rm ext})} \big( \gamma - 1/2 \big) \Big) + \frac{e^2}{2} \int_{\R^3} \int_{\R^3} \frac{\rho_{\gamma - 1/2}(x) \rho_{\gamma - 1/2}(y)}{|x - y|} \, dx \, dy\\
& - \frac{e^2}{2} \int_{\R^3} \int_{\R^3} \frac{|(\gamma - 1/2)(x, y)|^2}{|x - y|} \, dx \, dy + \frac{1}{8 \pi} \int_{\R^3}|\curl A(x)|^2 \, dx.
\end{align*}
The derivation of the (reduced) Bogoliubov-Dirac-Fock model from this energy consists in omitting some terms on one hand, introducing a regularization on the other hand. The regularization is based on the introduction of an ultraviolet cut-off $\Lambda$ in the Fourier space. This choice breaks the magnetic gauge invariance corresponding to replace $A$ by $A + \nabla \varphi$, which is used in Quantum Electrodynamics to eliminate some divergences in the perturbative computations. As a result, a relevant model including photons and external magnetic fields cannot rely on a sharp ultraviolet cut-off.

In order to derive an alternative model, it is convenient to express the direct electrostatic term in \eqref{eq:HF-relativistic} in terms of the Coulomb potential
$$V_{\gamma - 1/2}(x) = e \int_{\R^3} \frac{\rho_{\gamma - 1/2}(y)}{|x - y|} \, dy,$$
according to the formula
$$\frac{e^2}{2} \int_{\R^3} \int_{\R^3} \frac{\rho_{\gamma - 1/2}(x) \rho_{\gamma - 1/2}(y)}{|x - y|} \, dx \, dy = e \int_{\R^3} \rho_{\gamma - 1/2}(x) \, V_{\gamma - 1/2}(x) \, dx - \frac{1}{8 \pi} \int_{\R^3} |\nabla V_{\gamma - 1/2}(x)|^2 \, dx.$$
Since the potential $V_{\gamma - 1/2}$ solves the maximization problem
$$e \int_{\R^3} \rho_{\gamma - 1/2}(x) \, V_{\gamma - 1/2}(x) \, dx - \frac{1}{8 \pi} \int_{\R^3} |\nabla V_{\gamma - 1/2}(x)|^2 \, dx = \sup_V \, \bigg\{ e \int_{\R^3} \rho_{\gamma - 1/2} \, V - \frac{1}{8 \pi} \int_{\R^3} |\nabla V|^2 \bigg\},$$
we can write the formal energy in \eqref{eq:HF-relativistic} as
$$\boE_{HF}^{\bA_{\rm ext}}(\gamma, A) = \sup_V \, \boL_{HF}^{\bA_{\rm ext}}(\gamma, \bA).$$
where we have set $\bA = (V, A)$. In this formula, the formal Hartree-Fock Lagrangian $\boL_{HF}^{\bA_{\rm ext}}$ is given by
\begin{align*}
\boL_{HF}^{\bA_{\rm ext}}(\gamma, \bA) = & \tr \big( D_{m, e (\bA + \bA_{\rm ext})} (\gamma - 1/2) \big) - \frac{e^2}{2} \int_{\R^3} \int_{\R^3} \frac{|(\gamma - 1/2)(x, y)|^2}{|x - y|} \, dx \, dy\\
& + \frac{1}{8 \pi} \int_{\R^3} \big( |\curl A(x)|^2 - |\nabla V(x)|^2 \big) \, dx.
\end{align*}
We can also introduce a reduced Lagrangian omitting the exchange electrostatic term according to the formula
\begin{align*}
\boL_{rHF}^{\bA_{\rm ext}}(\gamma, \bA) = & \tr \big( D_{m, e (\bA + \bA_{\rm ext})} (\gamma - 1/2) \big) + \frac{1}{8 \pi} \int_{\R^3} \big( |\curl A(x)|^2 - |\nabla V(x)|^2 \big) \, dx.
\end{align*}
In the sequel, we restrict our attention to the reduced formalism. To our knowledge, the analysis of the original model remains largely open (see \cite{GrHaLeS1} for more details).

The polarized vacuum is constructed as a minimizer of the formal reduced energy
$$\boE_{rHF}^{\bA_{\rm ext}}(\gamma, A) = \sup_V \, \boL_{rHF}^{\bA_{\rm ext}}(\gamma, \bA),$$
with respect to the one-body density matrix $\gamma$ and the classical photon field $A$. The operator $\gamma$ is an orthogonal projector as before, while the field $A$ is divergence free due to the Coulomb gauge. In case of existence, a minimizer $(\gamma_*, A_*)$ is solution to the self-consistent equations
\begin{equation}
\label{eq:rHF-bA}
\begin{cases}
\gamma_* = \chi_{(- \infty, 0]} \big( D_{m, e(\bA_* + \bA_{\rm ext})} \big),\\
- \Delta A_* = 4 \pi \, e \, j_{\gamma_* - 1/2},\\
- \Delta V_* = 4 \pi \, e \, \rho_{\gamma_* - 1/2},\\
\div A_* = \div A_{\rm ext} = 0.
\end{cases}
\end{equation}
In this expression, the charge density $\rho_{\gamma_* - 1/2}$ and the charge current $j_{\gamma_* - 1/2}$ are given by
$$\rho_{\gamma_ * - 1/2}(x) = \tr_{\C^4} \big( (\gamma_* - 1/2) (x, x) \big) \quad {\rm and} \quad j_{\gamma_* - 1/2}(x) = \tr_{\C^4} \big( \bsalpha (\gamma_* - 1/2)(x, x) \big),$$
where $(\gamma_* - 1/2)(x, y)$ refers as above to the (formal) kernel of the operator $\gamma_* - 1/2$. The first equation in \eqref{eq:rHF-bA} is consistent with the picture of the Dirac sea since the minimizer $\gamma_*$ is the negative spectral projector of the Fock operator $D_{m, e(\bA_* + \bA_{\rm ext})}$. The equations for the electromagnetic four-potential $\bA_* = (V_*, A_*)$ are well-known in the Physics literature (see e.g. \cite{Engel1}).

Regarding the construction of electronic structures with $N$ electrons, a charge constraint of the form
$$\tr \big( \gamma - 1/2 \big) = N,$$
is added as for the reduced Bogoliubov-Dirac-Fock model. The equations for the minimizers $(\gamma_ N, A_N)$ write as
$$\begin{cases}
\gamma_N = \chi_{(- \infty, \mu_N]} \big( D_{m, e(\bA_N + \bA_{\rm ext})} \big),\\
- \Delta A_N = 4 \pi \, e \, j_{\gamma_N - 1/2},\\
- \Delta V_N = 4 \pi \, e \, \rho_{\gamma_N - 1/2},\\
\div A_N = \div A_{\rm ext} = 0,
\end{cases}$$
where $\mu_N$ is the Fermi level of the electronic structure. According to the picture of the Dirac sea, the physical electrons are identified to the projector $\gamma_N^{\rm ph} = \chi_{(0, \mu_N]}(D_{m, e(\bA_N + \bA_{\rm ext})})$, while the virtual electrons of the polarized vacuum are represented by $\gamma_N^{\rm vac} = \chi_{(- \infty, 0]}(D_{m, e(\bA_N + \bA_{\rm ext})})$. The model describes a positronic structure with $|N|$ positrons when $N$ is a negative integer. In the sequel, we focus on the construction of the polarized vacuum. Constructing electronic or positronic structures also remains an open problem.

Instead of maximizing the Lagrangian $\boL_{rHF}^{\bA_{\rm ext}}(\gamma, \bA)$ with respect to $V$ and then minimizing the resulting quantity with respect to $\gamma$ and $A$, one can follow the alternative strategy which consists in minimizing first with respect to $\gamma$, and then looking for a saddle point with respect to $V$ and $A$. We do not claim that we are solving exactly the same problem in this way, but this alternative strategy provides a polarized vacuum which is consistent with the self-consistent equations \eqref{eq:rHF-bA} (see Theorem \ref{thm:polarized}). Moreover, the problem is simplified. The unique minimizer of the minimization problem
\begin{equation}
\label{eq:min-gamma}
\inf_\gamma \, \boL_{rHF}^{\bA_{\rm ext}}(\gamma, \bA),
\end{equation}
is indeed explicitly given by
$$\gamma_\bA = \chi_{(- \infty, 0]} \big( D_{m, e(\bA + \bA_{\rm ext})} \big),$$
with a value for the minimum equal to
$$\inf_\gamma \, \boL_{rHF}^{\bA_{\rm ext}}(\gamma, \bA) = - \frac{1}{2} \tr \big| D_{m, e (\bA + \bA_{\rm ext})} \big| + \frac{1}{8 \pi} \int_{\R^3} \big( |\curl A|^2 - |\nabla V|^2 \big).$$
In order to construct the polarized vacuum, it only remains to solve a min-max problem which only depends on $V$ and $A$, namely
$$\min_A \, \max_V \, \Big\{ - \frac{1}{2} \tr \big| D_{m, e (\bA + \bA_{\rm ext})} \big| + \frac{1}{8 \pi} \int_{\R^3} \big( |\curl A|^2 - |\nabla V|^2 \big) \Big\},$$
where $|T|$ stands for the absolute value of the operator $T$.

At this stage, it is necessary to acknowledge that most of the previous discussion is only formal. We now have to introduce regularization techniques from Quantum Electrodynamics in order to define a rigorous model. The first element in this direction is reminiscent from the (reduced) Bogoliubov-Dirac-Fock model. When $\bA = \bA_{\rm ext} = 0$, the negative spectral projector $P_{m, 0}^-$ of the free Dirac operator formally solves the problem \eqref{eq:min-gamma}. However, the value of the minimum is equal to
$$\inf_\gamma \, \boL_{rHF}^0(\gamma, 0) = - \frac{1}{2} \tr \big| D_{m, 0} \big|,$$
which is an infinite quantity. In order to deal with finite quantities, and since the situation under our consideration corresponds to the free vacuum, we define a relative energy according to the formula
$$\boE_{\rm rel}^{\bA_{\rm ext}}(\bA) = \frac{1}{2} \tr \Big( \big| D_{m, 0} \big| - \big| D_{m, e (\bA + \bA_{\rm ext})} \big| \Big) + \frac{1}{8 \pi} \int_{\R^3} \big( |\curl A|^2 - |\nabla V|^2 \big),$$
so that the free vacuum now has a vanishing energy. Since this amounts to adding an infinite constant, this does not change the variational problem under our analysis, but we can now hope that the quantity $\boE_{\rm rel}^{\bA_{\rm ext}}(\bA)$ is finite provided $\bA$ and $\bA_{\rm ext}$ belong to some suitable function space. 

Actually, the energy remains divergent for large momenta. As a matter of fact, the operator $|D_{m, 0}| - |D_{m, e (\bA + \bA_{\rm ext})}|$ is not trace class when $A + A_{\rm ext} \neq 0$, so that its trace is not well-defined (see \cite{NencSch1} for more details). In order to remove these divergences, an ultraviolet cut-off is necessary. Various techniques from Quantum Electrodynamics are available. Our choice is to rely on a regularization proposed by Pauli and Villars in \cite{PaulVil1}, which consists in introducing the functional
\begin{equation}
\label{def:E-PV}
\boE_{\rm PV}^{\bA_{\rm ext}}(\bA) = \frac{1}{2} \tr \Bigg( \sum_{j = 0}^J c_j \, \Big( \big| D_{m_j, 0} \big| - \big| D_{m_j, e (\bA + \bA_{\rm ext})} \big| \Big) \Bigg) + \frac{1}{8 \pi} \int_{\R^3} \big( |\curl A|^2 - |\nabla V|^2 \big).
\end{equation}
In this expression, the index $j = 0$ corresponds to the physical electron-positron field. In particular, $m_0 = m$ is the (bare) mass of the electron. The indices $j = 1$ and $j = 2$ describe fictitious heavy particle fields. Their role is to remove the worst ultraviolet divergences (the linear ones to be more precise). In order to reach this goal, it is well-known (see \cite{PaulVil1}) that the coefficients $c_j$ and the masses $m_j$ must fulfil the two conditions
\begin{equation}
\label{eq:cond-PV}
\sum_{j = 0}^J c_j = \sum_{j = 0}^J c_j m_j^2 = 0.
\end{equation}
At least two additional distinct masses $m_1$ and $m_2$ are therefore necessary. In the sequel, we fix $J = 2$ and $c_0 = 1$. For this choice, the condition \eqref{eq:cond-PV} is equivalent to
$$c_1 = \frac{m_0^2 - m_2^2}{m_2^2 - m_1^2} \quad {\rm and} \quad c_2 = \frac{m_1^2 - m_0^2}{m_2^2 - m_1^2}.$$

In the limit $m_1 \to + \infty$ and $m_2 \to + \infty$, the regularization does not prevent a logarithmic divergence which is identical to the divergence of the constant $B_\Lambda$ in \eqref{eq:dev-B-Lambda} (see Proposition \ref{prop:def-F}). The divergence is best understood in terms of the averaged ultraviolet cut-off $\Lambda$ defined as
\begin{equation}
\label{def:Lambda}
\log(\Lambda^2) = - \sum_{j = 0}^2 c_j \log (m_j^2).
\end{equation}
The value of $\Lambda$ does not uniquely determine the masses $m_1$ and $m_2$. In practice, they are chosen as functions of $\Lambda$ such that the coefficients $c_1$ and $c_2$ remain bounded when $\Lambda \to + \infty$.

In the Coulomb gauge, the Pauli-Villars regulated energy $\boE_{\rm PV}^{\bA_{\rm ext}}$ is rigorously well-defined under the natural conditions that the fields $B = \curl A$, $B_{\rm ext} = \curl A_{\rm ext}$, $E = - \nabla V$ and $E_{\rm ext} = - \nabla V_{\rm ext}$ are square integrable (see Proposition \ref{prop:def-F}). The polarized vacuum is described using a solution to the min-max problem
\begin{equation}
\label{def:mini-PV}
\min_A \, \max_V \, \boE_{\rm PV}^{\bA_{\rm ext}}(\bA).
\end{equation}
More precisely, it is identified to the projector
$$\gamma_{\rm PV}^{\rm vac} = \chi_{(- \infty, 0]} \big( D_{m_0, e(\bA_* + \bA_{\rm ext})} \big)$$
where $\bA_* = (V_*, A_*)$ is a saddle point for the problem \eqref{def:mini-PV}. One can guarantee the existence of such a saddle point at least when the external electromagnetic field $\bF_{\rm ext} = (E_{\rm ext}, B_{\rm ext})$ is weak enough.

\subsection{Construction of the polarized vacuum}
\label{sub:const-PV}

We now define properly the Pauli-Villars regulated energy in \eqref{def:E-PV} before solving the min-max problem \eqref{def:mini-PV}. The natural framework for defining the energy $\boE_{\rm PV}^{\bA_{\rm ext}}$ is provided by the Coulomb-gauge homogeneous Sobolev space
$$\dot{H}_{\rm div}^1(\R^3) = \Big\{ \bA = (V, A) \in L^6(\R^3, \R^4), \ {\rm s.t.} \ \div A = 0 \ {\rm and} \ \bF=(-\nabla V, \curl A) \in L^2(\R^3, \R^6) \Big\},$$
which is an Hilbert space for the norm
$$\| \bA \|_{\dot{H}_{\rm div}^1(\R^3)}^2 = \| \nabla V \|_{L^2(\R^3)}^2 + \| \curl A \|_{L^2(\R^3)}^2 = \| \bF \|_{L^2(\R^3)}^2.$$
When $\bA \in \dot{H}_{\rm div}^1(\R^3)$, the integral in \eqref{def:E-PV} is well-defined, but we have to provide a rigorous meaning to the first term in \eqref{def:E-PV}. This amounts to defining properly the functional 
\begin{equation}
\label{eq:formal-F}
\boF_{\rm PV}\big( \bA \big) = \frac{1}{2} \tr \sum_{j = 0}^2 c_j \Big( \big| D_{m_j, 0} \big| - \big| D_{m_j, \bA } \big| \Big),
\end{equation}
for an arbitrary four-potential $\bA \in \dot{H}_{\rm div}^1(\R^3)$. In this direction, we can establish the next proposition.

\begin{prop}[\cite{GrHaLeS1}]
\label{prop:def-F}
Assume that the coefficients $c_j$ and the masses $m_j$ satisfy
\begin{equation}
\label{eq:cond-cj-mj}
c_0 = 1, \quad m_2 > m_1 > m_0 > 0 \quad {\rm and} \quad \sum_{j = 0}^2 c_j = \sum_{j = 0}^2 c_j m_j^2 = 0.
\end{equation}
$(i)$ Let
\begin{equation}
\label{def:TA}
T_\bA = \frac{1}{2} \sum_{j = 0}^2 c_j \Big( \big| D_{m_j, 0} \big| - \big| D_{m_j, \bA} \big| \Big).
\end{equation}
Given any $\bA \in L^1(\R^3) \cap \dot{H}_{\rm div}^1(\R^3)$, the operator $\tr_{\C^4} T_\bA$ is trace class on $L^2(\R^3, \C)$. In particular, the quantity $\boF_{\rm PV}(\bA)$ is well-defined by the expression
\begin{equation}
\label{def:F}
\boF_{\rm PV}(\bA) = \tr \big( \tr_{\C^4} T_{\bA} \big).
\end{equation}
$(ii)$ The functional $\boF_{\rm PV}$ can be uniquely extended to a continuous mapping on $\dot{H}_{\rm div}^1(\R^3)$.\\
$(iii)$ Let $\bA \in \dot{H}_{\rm div}^1(\R^3)$. We have
\begin{equation}
\label{eq:devF}
\boF_{\rm PV}(\bA) = \boF_2(\bF) + \boR(\bA),
\end{equation}
where $\bF = (E, B)$, with $E = - \nabla V$ and $B = \curl A$. The functional $\boR$ is continuous on $\dot{H}_{\rm div}^1(\R^3)$ and satisfies
\begin{equation}
\label{est:R}
|\boR(\bA)| \leq K \Bigg( \bigg( \sum_{j = 0}^2 \frac{|c_j|}{m_j} \bigg) \big\| \bF \big\|_{L^2}^4 + \bigg( \sum_{j = 0}^2 \frac{|c_j|}{m_j^2} \bigg) \big\| \bF \big\|_{L^2}^6 \Bigg),
\end{equation}
for a universal positive number $K$.\\
$(iv)$ The functional $\boF_2$ is the non-negative and bounded quadratic form on $L^2(\R^3, \R^4)$ given by 
\begin{equation}
\label{eq:F2}
\boF_2(\bF) = \frac{1}{8 \pi} \int_{\R^3} M(k) \Big( \big| \widehat{B}(k) \big|^2 - \big| \widehat{E}(k) \big|^2 \Big) \, dk, 
\end{equation}
where
\begin{equation}
\label{def:M}
M(k) = - \frac{2}{\pi} \sum_{j = 0}^2 c_j \int_0^1 u (1 - u) \log \big( m_j^2 + u (1 - u) |k|^2 \big) \, du.
\end{equation}
The function $M$ is positive and satisfies the uniform estimate
\begin{equation}
\label{est:M}
0 < M(k) \leq M(0) = \frac{2 \log(\Lambda)}{3 \pi},
\end{equation}
where $\Lambda$ is given by \eqref{def:Lambda}.
\end{prop}

The proof of Proposition \ref{prop:def-F} relies on a perturbative expansion of the operator $T_\bA$. Invoking the formula
$$|T| = \frac{1}{2 \pi} \int_\R \Big( 2 - \frac{i \omega}{T + i \omega} + \frac{i \omega}{T - i \omega} \Big) \, d\omega,$$
the operator $T_\bA$ may be written as
$$T_\bA = \frac{1}{4 \pi} \int_\R \sum_{j = 0}^2 c_j \, \Big( \frac{i \omega}{D_{m_j, \bA} + i \omega} - \frac{i \omega}{D_{m_j, \bA} - i \omega} - \frac{i \omega}{D_{m_j, 0} + i \omega} + \frac{i \omega}{D_{m_j, 0} - i \omega} \Big) \, d\omega.$$
Expanding with respect to the powers of $\bA$ leads to the expression
\begin{align*}
T_\bA = & \sum_{n = 1}^5 T_n(\bA) + T_6'(\bA)\\
= & \frac{1}{4 \pi} \sum_{n = 1}^5 \int_\R \big( R_n(\omega, \bA) + R_n(- \omega, \bA) \big) \, d\omega + \frac{1}{4 \pi} \int_\R \big( R_6'(\omega, \bA) + R_6'(- \omega, \bA) \big) \, d\omega,
\end{align*}
with
$$R_n(\omega, \bA) = \sum_{j = 0}^2 c_j \, \frac{i \omega}{D_{m_j, 0} + i \omega} \Big( \big( \bsalpha \cdot A - V \big) \frac{1}{D_{m_j, 0} + i \omega} \Big)^n,$$
for $1 \leq n \leq 5$, and
$$R_6'(\omega, \bA) = \sum_{j = 0}^2 c_j \, \frac{i \omega}{D_{m_j, \bA} + i \omega} \Big( \big( \bsalpha \cdot A - V \big) \frac{1}{D_{m_j, 0} + i \omega} \Big)^6.$$
Due to the conditions \eqref{eq:cond-cj-mj}, the operators $\tr_{\C^4} T_n(\bA)$ and $\tr_{\C^4} T_6'(\bA)$ are trace class on $L^2(\R^3)$ when $\bA$ belongs to $L^1(\R^3) \cap \dot{H}_{\rm div}^1(\R^3)$. The quantity $\boF_{\rm PV}(\bA)$ in \eqref{def:F} is therefore well-defined. Moreover, it depends H\"older continuously on $\bA \in \dot{H}_{\rm div}^1(\R^3)$, so that it can be extended to the space $\dot{H}_{\rm div}^1(\R^3)$.

Let us emphasize the introduction of the $\C^4$-trace here. The operators $T_n(\bA)$ and $T_6'(\bA)$ are probably not trace class without taking first the $\C^4$-trace (except when $A = 0$). Defining $\boF_{\rm PV}$ as in \eqref{def:F} extends the formal definition \eqref{eq:formal-F} to the case where $T_\bA$ is not a trace-class operator. The two definitions remain identical when $T_\bA$ is trace class.

Concerning the second order operator $\tr_{\C^4} T_2(\bA)$, an explicit computation leads to the formula
$$\tr \big( \tr_{\C^4} T_2(\bA) \big) = \boF_2(\bF),$$
where $\boF_2(\bF)$ is defined in \eqref{eq:F2}. The Fourier multiplier $M$ in \eqref{eq:F2} describes the linear response of the virtual electrons in the polarized vacuum. In view of the convergence
\begin{equation}
\label{eq:lim-M}
\lim_{\Lambda \to \infty} \Big( \frac{2 \log \Lambda}{3 \pi} - M(k) \Big) = U(k),
\end{equation}
where $U$ is the Uehling multiplier given by \eqref{def:Uehling}, the function $M$ appears as the Pauli-Villars equivalent of the function $B_\Lambda$ in the reduced Bogoliubov-Dirac-Fock model. This similarity in the two models results from the gauge and relativistic invariances in Quantum Electrodynamics.

It follows from \eqref{eq:lim-M} that the self-consistent equations of the charge and current densities corresponding to a (possible) solution $\bA_*$ to the min-max problem \eqref{def:mini-PV} are very similar to the equation \eqref{eq:rho-Q*} for the charge density $\rho_{Q_*}$ of a minimizer $Q_*$ of the reduced Bogoliubov-Dirac-Fock model. Even if this was not done in \cite{GrHaLeS1}, the renormalization technique applied to define a physical charge density $\rho_{\rm ph}$ and to compute its perturbative expansion with respect to $\alpha_{\rm ph}$ in the case of the reduced Bogoliubov-Dirac-Fock model is likely to work the same with the Pauli-Villars regulated model corresponding to the energy $\boE_{\rm PV}^{\bA_{\rm ext}}$.

This however requires to construct first a solution $\bA_*$ to the min-max problem \eqref{def:mini-PV}. The construction is possible when the external electromagnetic four-potential $\bA_{\rm ext}$ is small enough. In this case, one can deduce from the expression of the second-order functional $\boF_2$ in \eqref{eq:F2} that the energy $\boE_{\rm PV}^{\bA_{\rm ext}}$ owns a local saddle point geometry close to the four-potential $\bA = 0$. The existence of a (local) solution to the min-max problem \eqref{def:mini-PV} follows using tools from convex analysis.

\begin{thm}[\cite{GrHaLeS1}]
\label{thm:polarized}
Assume that the coefficients $c_j$ and the masses $m_j$ satisfy the conditions \eqref{eq:cond-cj-mj}.\\
$(i)$ There exists a positive number $r$ such that, given any four-potential $\bA_{\rm ext} \in \dot{H}_{\rm div}^1(\R^3)$ such that
\begin{equation}
\label{eq:estim_A_ext}
e \| \bA_{\rm ext} \|_{\dot{H}_{\rm div}^1(\R^3)} < \frac{r \sqrt{m_0}}{8},
\end{equation}
there exists a unique solution $\bA_* \in \dot{H}_{\rm div}^1(\R^3)$ to the min-max problem 
\begin{equation}
\label{eq:maxmin_minmax}
\begin{split}
\boE_{\rm PV}^{\bA_{\rm ext}}(\bA_*) & = \max_{\| \nabla V \|_{L^2} < \frac{r \sqrt{m_0}}{4 e}} \quad \inf_{\| \curl A \|_{L^2} < \frac{r \sqrt{m_0}}{4 e}} \, \boE_{\rm PV}^{\bA_{\rm ext}}(\bA)\\
& = \min_{\| \curl A \|_{L^2} < \frac{r \sqrt{m_0}}{4 e}} \quad \sup_{\| \nabla V \|_{L^2} < \frac{r \sqrt{m_0}}{4 e}} \, \boE_{\rm PV}^{\bA_{\rm ext}}(\bA).
\end{split}
\end{equation}
$(ii)$ When $\bA_{\rm ext} = 0$, the solution $\bA_*$ is equal to $0$.\\
$(iii)$ The four-potential $\bA_*$ is a solution to the nonlinear equations
\begin{equation}
\label{eq:SCF}
\Bigg\{ \begin{array}{ll} - \Delta V_* = 4 \pi e\,\rho_*,\\
- \Delta A_* = 4 \pi e\, j_*, \end{array}
\end{equation}
where $\rho_* \in \boC(\R^3)$ and $j_* \in \boC(\R^3)$ are defined as
\begin{equation}
\label{eq:def-rho-j*}
\rho_*(x) = \big[ \tr_{\C^4} Q_* \big](x, x) \quad {\rm and} \quad j_\bA(x) = \big[ \tr_{\C^4} \bsalpha \, Q_* \big](x, x).
\end{equation}
In this expression, the function $Q_*(x, y)$ refers to the kernel of the locally trace-class operator
\begin{equation}
\label{eq:SCF_op}
Q_* = \sum_{j = 0}^2 c_j \, \chi_{(- \infty, 0]} \big( D_{m_j, e (\bA_* + \bA_{\rm ext})} \big).
\end{equation}
\end{thm}

According to the previous derivation of the Pauli-Villars regulated energy, the polarized vacuum is identified to the projector
$$\gamma_{\rm PV}^{\rm vac} = \chi_{(- \infty, 0]} \big( D_{m_0, e (\bA_* + \bA_{\rm ext})} \big).$$
Its construction is only local and only available for small enough external electromagnetic fields.

To our knowledge, the existence of a global solution to the min-max problem \eqref{def:mini-PV} remains an open problem. A first attempt to answer this question could concern the property that $\bA = 0$ is the unique global saddle point of the energy $\boE_{\rm PV}^0$.

The construction of a (local) minimizer in large external electromagnetic fields is another appealing problem, in particular since it certainly requires to understand the phenomenon of production of electron-positron pairs (see \cite{Sabin1} for a first analysis of this phenomenon).

\bibliographystyle{plain}
\bibliography{Bibliogr}

\end{document}